\documentclass{aa}
\usepackage{graphicx}
\usepackage{natbib}
\bibpunct{(}{)}{;}{a}{}{,} %

\newcommand{\ergs}{\ensuremath{\mathrm{erg\,s}^{-1}}}
\newcommand{\ergscm}{\ensuremath{\mathrm{erg\,s}^{-1}\,\mathrm{cm}^{-2}}}
\newcommand{\ergshz}{\ensuremath{\mathrm{erg\,s}^{-1}\,\mathrm{Hz}^{-1}}}
\newcommand{\ergscma}{\ensuremath{\mathrm{erg\,s}^{-1}\,\mathrm{cm}^{-2}\,\mathrm{\AA}^{-1}}}

\newcommand{\ergscmhz}{\ensuremath{\mathrm{erg\,s}^{-1}\,\mathrm{cm}^{-2}\,\mathrm{Hz}^{-1}}}

\begin{document}
\title{Probing the nuclear obscuration in radio-galaxies with near
  infrared imaging\thanks{Based on observations made with the 
  Italian Telescopio Nazionale Galileo (TNG) operated on the island 
  of La Palma by the Centro Galileo Galilei of the INAF (Istituto 
  Nazionale di Astrofisica) at the Spanish Observatorio del Roque 
  de los Muchachos of the Instituto de Astrofisica de 
  Canarias.}\fnmsep\thanks{Based on observation obtained at the Space 
  Telescope Science Institute, which is operated by the Association of 
  Universities for Research in Astronomy, Incorporated, under NASA 
  contract NSA 5-26555.}}

\author{D. Marchesini\inst{1}\fnmsep\thanks{\emph{Present address:} Department of Astronomy, Yale University, P.O. Box 208101, New Haven, CT 06520-8101, U.S.A.} \and A. Capetti\inst{2} \and A. Celotti\inst{1}}

  \offprints{D. Marchesini, \email{danilom@astro.yale.edu}}

\institute{
  SISSA/ISAS, Via Beirut 2-4, I-34014 Trieste, Italy \\
  \and
  INAF - Osservatorio Astronomico di Torino, Strada Osservatorio 20,
  I-10025 Pino Torinese, Italy} 

\date{Received ...; accepted 08 December 2004}

\abstract{We present the first near-infrared ($K^{\prime}$-band)
  homogeneous observations of a complete sub-sample of the 3CR radio
  catalogue comprising all High Excitation Galaxies (HEGs) at $z <$0.3.
  After showing that the surface brightness decomposition technique to
  measure central point-like sources is affected by significant
  uncertainties for the objects in the studied sample, we present a
  new, more accurate method based on the $R-K^{\prime}$ color profile.
  Via this method we find a substantial nuclear $K^{\prime}$-band excess in all
  but two HEGs -- most likely directly associated to their nuclear
  emission -- and we measure the corresponding 2.12~$\mu$m nuclear
  luminosities.  Within the frame of the unification scheme for
  radio-loud active galactic nuclei, it appears that obscuration alone
  is not able to account for the different nuclear properties of the
  majority of the HEGs and Broad Line Radio Galaxies (BLRGs), and also
  scattering of the (optically) hidden nuclear light from a compact
  region must be invoked.  More precisely, for $\sim$70\% of the HEGs
  the observed point-like optical emission is dominated by the
  scattered component, while in the $K^{\prime}$-band both scattered
  and direct light passing through the torus contribute to the
  observed nuclear luminosity.  The estimated fraction of scattered
  light ranges from a few tenths to a few percent, while the torus
  extinction is between 15$<A_{\rm V,~torus}<$50~mag with only a few
  exceptions with lower obscuration.
  
  }

\authorrunning{Marchesini et al.}  \titlerunning{Probing obscuration
  in radio-galaxies} \maketitle


\section{Introduction}

Unification models for radio-loud active galactic nuclei (RL AGNs) aim
at understanding the different observed properties of intrinsically
identical sources in terms of anisotropies in the nuclear emission
(see \citealt{urry95} for a review). Anisotropy can arise from 
relativistic beaming and/or selective obscuration caused by material
arranged in a non-spherical, torus-like geometry.

RL AGNs can be broadly divided into two classes based on their
extended radio morphology: low-power, edge-darkened FR~I and
high-power, edge-brightened FR~II sources \citep{fanaroff74}. For the
powerful FR~II radio-galaxies (RGs), the unification of the
broad-lined (BLRGs) and narrow-lined objects (NLRGs) invokes the
existence of an obscuring `torus' blocking the direct view of the
nuclear and broad-line emission at large angles with respect to the
jet axis.  The detection of polarized broad emission lines in (a still
small number of) NLRGs (e.g. \citealt{antonucci84}; \citealt{cohen99};
\citealt{young96a}; \citealt{ogle97}) provides the most direct support
to this scenario. Conversely, in low-luminosity FR~I RGs obscuration
is actually neither `required' nor supported observationally
\citep{urry95}.

A complication in this scheme arises due to the spectral inhomogeneity
of the FR~II RG population, comprising within the NLRGs class both
high-excitation (HEGs) and low-excitation (LEGs) galaxies
(\citealt{laing94}; \citealt{jackson97}), while all FR~Is are
characterized by low-excitation spectra.

The study of the nuclear properties of RGs strongly benefited from the
results of Hubble Space Telescope (HST) optical and UV imaging of most
3CR RGs (\citealt{spinrad85}) with redshift $z<$0.3 (33 FR~Is and 63
FR~II; Chiaberge, Capetti \& Celotti 1999, 2002, CCC99, CCC02
hereafter; Chiaberge et al. 2002).  Unresolved nuclear sources are
detected in most of the FR~Is, suggesting that their nuclei are
essentially unobscured (CCC99). The observed large range in the
optical-UV spectral index can be accounted for just by moderate
absorption ($A_{\rm V} \sim$1-2~mag), due to either extended
(kpc-scale) dust lanes or ($\sim$100~pc) dusty disks
\citep{chiaberge02}. Furthermore, the nuclear optical luminosities
strongly correlate with the 5~GHz radio core ones, arguing for a
common synchrotron origin, and in turn correspond to extremely low
Eddington ratios (e.g. \citealt{fabian95}; \citealt{zirbel95};
\citealt{willott00}; \citealt{wills04}). Thus FR~I RGs might be
lacking a substantial broad line region, torus and thermal disc
emission.  Unresolved nuclear components are also detected in a
significant number of FR~IIs (CCC02). With respect to FR~Is they show
a more complex behaviour which is however clearly related to their
spectral classification: i) BLRGs have optical cores with luminosities
in excess to those of FR~I of similar radio power, possibly dominated
by thermal emission; ii) no central source is seen in several HEGs,
likely constituting the obscured narrow-lined counterparts of BLRGs;
iii) in a few objects (e.g. 3CR~109 and 3CR~234) the detected nuclear
component is consistent with being the (obscured) quasar light
scattered into our line-of-sight; iv) LEGs harbor faint optical cores,
essentially indistinguishable from those seen in FR~Is, indicating
that the FR~I/II dichotomy is not univocally connected with the
nuclear structure (LEGs constitute a substantial fraction of FR~II
sources).

Nonetheless, the optical results can be strongly hampered by the
effects of dust extinction. This problem is certainly serious in RGs
where (even leaving aside the presence of nuclear tori) nuclear discs
and dust lanes do not allow us to study the central regions in a
significant fraction of the 3CR sources ($\sim$ 30 \%)
\citep{dekoff00}.  Particularly instructive is the case of Centaurus
A, the nearest RG.  An unresolved nuclear component is seen in HST
images from the $V$ to the $K$ band \citep{marconietal00} but its
infrared flux is five orders of magnitude larger than the optical one.
A similar effect is also commonly seen in Seyfert 2, where HST imaging
have revealed the ubiquitous presence of near-infrared (NIR) nuclei in
optically obscured sources (e.g.  \citealt{kulkarni98};
\citealt{quillen01}).  Thus to gain a better understanding of the
physics of RL AGNs it is crucial to obtain a more unbiased view of RGs
nuclei, which can be provided by NIR observations.

Relatively little effort has been devoted to date to NIR imaging of
RGs (\citealt{lilly82}; \citealt{taylor96}; \citealt{simpson00}, 
hereafter SWW00) and no systematic program has been carried out.
Moreover in \citet{lilly82} only 10\arcsec-12\arcsec~aperture
photometry has been performed, leading to only about a dozen objects
with NIR nuclear luminosity estimates, out of the complete 3CR RG
sample with $z<$0.3.  NIR spectroscopy \citep{hill96} and thermal-IR
($L^{\prime}$- and $M$-band) imaging \citep{simpson99} for a total of
about 15 RGs have revealed quasar nuclei obscured by $A_{\rm V}$ from
$\sim 0-15$ to $>$45~mag (SWW00).

The NIR results can/will be also compared to those in the 
mid-infrared (MIR) and in the far-infrared (FIR). Indeed the recent
study of 3CR sources by \citet{haas04} provided evidence for the
presence in high-power RGs of powerful AGN hidden by a dusty torus:
steep-spectrum radio quasar, BLRGs and HEGs showed very similar IR
spectral energy distributions, characterized by a FIR dust emission
component, with relatively weaker MIR, NIR and optical emission
in HEGs with respect to BLRGs. On the contrary, FR~Is and LEGs were
characterized by a different spectral energy distributions, showing a
dust emission bump in the MIR-FIR from dust cooler than in FR~II, a
bump peaking at 1~$\mu$m due to the elliptical hosts and a steep
synchrotron spectrum in the cm-mm wavelength range. A comparison
between 3CR FR~Is and high-power FR~IIs in the MIR and FIR has been
also recently performed by \citet{muller04} showing that FR~Is are
characterized by MIR and FIR dust luminosities much lower than FR~IIs,
consistent with the former hosting low-luminosity nuclei.

To extend the NIR diagnostic to a large complete sample we performed a
$K^{\prime}-$band survey of all RG belonging to the 3CR sample at $z
<$ 0.3, aiming at: i) identifying optically obscured sources from the
presence of a NIR excess with respect to the extrapolation of the
optical (HST) flux; ii) establishing the role of obscuration in FR~I
RGs; iii) exploring the nature of narrow lined FR~IIs, in particular
those with faint optical nuclei; iv) comparing the NIR properties of
low and high luminosity radio sources; v) studying the properties of
the nuclear regions of the sources affected by large scale dust
obscuration.

In this paper we present results obtained for the subsample of FR~II
RGs with an HEG optical spectrum.  The outline is the following. In
\S~\ref{observations} we describe the IR observations and data
reduction; a comparison between the IR and HST images aimed at finding
obscured nuclei is presented in \S~\ref{method}. In \S~\ref{results}
and \S~\ref{discussion} we focus on the results and discuss some of
their consequences.  Our conclusions are drawn in the final
\S~\ref{conclusion}. A zero cosmological constant universe with
$H_{0}$=75~km~s$^{-1}$~Mpc$^{-1}$ and $q_{0}$=0.5 is adopted
throughout the paper for comparison with previous works in this field.


\section{The sample, observations and data reduction}
\label{observations}

\subsection{The sample}

Among the 3CR RGs at $z<0.3$, the HEG sub--sample comprises 32 objects
(CCC02). Following the definition by \citet{jackson97}, HEGs have been
classified on the basis of their [OIII] equivalent width (larger than
10~\AA), and/or the [OII]/[OIII] ratio (smaller than 1). In
Table~\ref{tab1}, names, redshifts and radio data of the objects are
reported, together with the optical [OIII] line and core continuum
(HST) luminosities, as taken from the literature.  Six
objects of the sample have been excluded from the analysis: 3CR~63,
3CR~135 and 3CR~192 were not observed because of scheduling problems
or clouds, 3CR~153 had very bad seeing, 3CR~33 and 3CR~105 have no
available HST image (see \S~\ref{irlum}).

\begin{table*}
\caption{The sample of HEGs}
\small
\begin{tabular}{l c c c c c c } \hline
Source Name &  Redshift & Images$^{a}$    & $\log L_{\rm 178}$ &  $\log L_{\rm r}$ &   $\log L_{\rm [OIII]}$ & $\log L_{\rm O}$ \\
            &    $z$    &           &  \ergshz        & \ergshz     &    \ergs            & \ergshz    \\
            &           &           &                 &             &                     &            \\
\hline
3C~18       &   0.188   &  H,A      &  34.16     &   31.84      &       41.88    &     28.62    \\
3C~33       &   0.059   &  A        &  33.56     &   30.27      &       42.01    &    not obs.  \\
3C~63       &   0.175   &  H        &  34.08     &   30.97      &        --      &     28.29    \\
3C~79       &   0.256   &  H,N      &  34.54     &   31.19      &       42.43    &     28.33    \\
3C~98       &   0.030   &  H,N      &  32.79     &   29.25      &       40.87    &  $<$25.67    \\
3C~105      &   0.089   &  N        &  33.45     &   30.36      &       40.78    &     not obs. \\
3C~135      &   0.127   &  H        &  33.78     &   30.18      &        --      &     27.36    \\
3C~153      &   0.277   &  H,N      &  34.44     &   29.73      &        --      &  $<$26.66    \\
3C~171      &   0.238   &  H,N      &  34.42     &   30.36      &       42.50    &     26.50    \\
3C~184.1    &   0.118   &  H,N      &  33.61     &   30.25      &       42.19    &     28.17    \\
3C~192      &   0.060   &  H        &  33.22     &   29.73      &       41.60    &  $<$26.99    \\
3C~197.1    &   0.131   &  H,N      &  33.45     &   30.30      &       41.26$^b$ &     28.10    \\
3C~198      &   0.082   &  H,N      &  33.38     &    --        &       41.04    &     28.03    \\
3C~223      &   0.137   &  H,N      &  33.74     &   30.57      &       42.17    &  $<$27.27    \\
3C~223.1    &   0.108   &  H,N      &  33.31     &   30.24      &       41.65    &  $<$27.18    \\
3C~234      &   0.185   &  H,N      &  34.27     &   31.88      &       43.17    &     29.10    \\
3C~284      &   0.239   &  H,N      &  33.87     &   30.25      &       42.19    &   complex    \\
3C~285      &   0.079   &  H,N      &  32.87     &   29.93      &       40.73    &     25.65    \\ 
3C~300      &   0.270   &  H,N      &  34.45     &   31.07      &       42.16    &     27.79    \\
3C~321      &   0.096   &  H,A      &  33.32     &   30.78      &       42.15    &    complex   \\
3C~327      &   0.104   &  H,A      &  33.97     &   30.88      &       42.05    &  $<$26.55    \\
3C~349      &   0.205   &  H,A      &  34.07     &   31.18      &       41.42    &     28.16    \\
3C~357      &   0.167   &  H,A      &  33.67     &   30.48      &       41.76$^c$ &  $<$26.90    \\
3C~379.1    &   0.256   &  H,A      &  34.03     &    --        &        --      &  $<$27.21    \\
3C~381      &   0.161   &  H,A      &  33.91     &   30.48      &       42.37    &  $<$27.52    \\
3C~402      &   0.025   &  H,A      &  32.04     &   29.73      &        --      &     26.59    \\
3C~403      &   0.059   &  H,A      &  33.28     &   29.87      &       41.55    &     26.65    \\  
3C~405      &   0.056   &  H,A      &  35.73     &   32.20      &        --      &  complex?    \\
3C~436      &   0.215   &  H,A      &  34.18     &   31.21      &       41.52    &    complex   \\
3C~452      &   0.081   &  H,A      &  33.88     &   31.24      &       41.35$^d$ &  $<$26.89    \\
3C~456      &   0.233   &  H,A      &  34.21     &   31.39      &       42.59    &     28.50    \\
3C~460      &   0.268   &  H,A      &  34.10     &   31.39      &       41.67    &     27.60    \\
\hline
\end{tabular}

{$^{a}$Images coding: H=HST, A=Arnica, N=NICS;

$L_{\rm 178}$ is the total radio luminosity at 178~MHz; $L_{\rm r}$ 
is the nuclear radio luminosity at 5~GHz (references in CCC02); 
$L_{\rm [OIII]}$ is the luminosity of the [OIII]~5007~\AA~emission line 
from \citet{jackson97} and C. Willott's web page, 
{\ttfamily http://www-astro.physics.ox.ac.uk/$\sim$cjw/3crr/3crr.html}; 
except for $^b$3CR~197.1 (from the SDSS spectrum), $^c$3CR~357 
\citep{crawford88} and $^d$3CR~452 \citep{rawlings89}; $L_{\rm O}$ is 
the luminosity of the optical core as measured in CCC02.}

\label{tab1}

\end{table*}

\subsection{TNG observations}

Observations have been obtained with the 3.6m Telescopio Nazionale
Galileo (TNG), the Italian national facility located at La Palma
Island (Spain) in two runs on Jul 8-12, 2000 and on Feb 9-13, 2001.
During the first run we used ARNICA (ARcetri Near Infrared CAmera), a
256$\times$256 pixel NICMOS 3 array with a pixel size of 0\farcs35\
and a field of view (FoV) of 1$^\prime$.5$\times$1$^\prime$.5, while
for the second run NICS (Near Infrared Camera Spectrometer), a
Rockwell 1024$\times$1024 HgCdTe Hawaii array, was already available.
In its small field mode the pixel size is $0\farcs13$ for a total FoV
of 2$^\prime$.2$\times$2$^\prime$.2, while in its large field mode
pixel size and FoV are doubled.

Images were taken with the dithering mode with the source located in a
pattern of 20 positions with a width of $\pm 30\arcsec$.  All sources
of this sub-sample are sufficiently small that it was possible to
always keep the source within the images during the pattern (for both
ARNICA and NICS).  For NICS, the individual integration time was set
to 1 min to achieve background limited images, i.e. a total
integration time of 20 min for each source (except 3CR~184.1 and
3CR~234, for which it was 16 and 15 min, respectively). For ARNICA the
individual integration time was set to 48 sec, and the total ranged
from 4.8 to 24 min depending on the source.  We used the $K^{\prime}$
filter that has a central wavelength of 2.12 $\mu$m and a FWHM of 0.35
$\mu$m.

\subsection{Data reduction}

The images taken with NICS were reduced with SNAP (Speedy
Near-infrared data Automatic Pipeline), a pipeline for automatic
reduction developed by F. Mannucci. SNAP performs a full reduction,
with flat-fielding, sky subtraction, computation of the offsets,
correction for geometrical distortion, object masking and correction
for cross-talk between the quadrants (for more information, see
{\ttfamily http://www.arcetri.astro.it/$\sim$filippo/snap/}).  For the
ARNICA data, a flat field was produced by median filtering of the
frames, after scaling each frame to have the same median pixel value.
Each frame was then divided by a normalized flat-field and registered
using as reference the target itself and, when possible, stars present
in the FoV. The aligned images were then averaged to produce the final
reduced image.

For both NICS and ARNICA, absolute flux calibration was obtained
observing 6-11 standard fields (depending on the night) using a 5
position dithering pattern. From each frame of each standard star we
derived a measurement of the zero-point, for a total of 32-61
independent values per night. The dispersion of the zero-point
measurements provides an estimate of the calibration accuracy in the
range 4-6\% for NICS and 3-6\% for ARNICA, depending on the night.
NICS last night was not photometric, with thin cirrus, and no standard
star has been observed. The absolute flux calibration was obtained
using the stars present in the fields in common with previous nights
(these fields were re-observed since the seeing in last night was much
better). The flux calibration proved to be stable throughout the
night, consistent with previous night calibrations and with an
accuracy of $\sim$ 6\%.

The absolute flux calibration has been compared with data published in
the literature, available for 3CR~79, 3CR~98, 3CR~171, 3CR~223,
3CR~234 (SWW00), 3CR~456 and 3CR~460 \citep{devries98}.  After
correcting for the different filter used
[$K^{\prime}=K+(0.22\pm0.03)(H-K)$; \citealt{neumann97}], our aperture
photometry is perfectly consistent with theirs (the mean difference is
0.02$\pm$0.03 mag, less than our absolute flux calibration error).
The images of all sources -- together with their corresponding HST ones, 
see \S~\ref{irlum} -- are shown in Appendix~\ref{figuresingle}.


\section{Analysis}
\label{method}

\subsection{The surface brightness decomposition technique} 
\label{stan_tech_analysis}

The `standard' technique used to detect and measure point sources at
galaxy centers requires a one (or two) dimensional synthetic modeling
of the galaxy profile to which a central point source is added.
Setting the seeing from observations, free parameters of this
procedure are those describing the galaxy profile (e.g.  central
surface brightness and effective radius) and the intensity of the
nuclear source. This technique works well with HST images, not
affected by seeing effects, and high quality ground-based photometry
of local sources.  Indeed only the HST spatial resolution and
sensitivity to high surface brightness features
(\citealt{hutchings95}; \citealt{mcleod95}) have allowed the detection
of faint point-like sources at the centre of typical RGs
(\citealt{capetti99}; CCC99), totally swamped by the host stellar
emission at ground-base resolution.

We have applied this procedure to measure the IR nuclear luminosity
$L_{\rm K^{\prime}}$ of our sources, by using the two-dimensional
fitting algorithm GALFIT \citep{peng02} designed to extract structural
components from galaxy images by simultaneously fit an arbitrary
number of them.  In Appendix~\ref{galfitApdx}, the results from
such analysis are presented and discussed for representative sources.
In summary we find that: 1) for the high $z$ sources and those with
faint nuclei (such as 3CR~405), the measured $L_{\rm K^{\prime}}$ can
differ by more than a factor of 20 depending on the initial parameters;
2) in many cases the fits are equally good with or without a central
point source [even in the case of 3CR~405, in which this is clearly
detected both in the HST image and with the $R-K^{\prime}$ color
profile analysis (see Appendix~\ref{colprofApdx})]; 3) consistent and
robust results are obtained only for those objects in which the
nuclear component clearly dominates over the host galaxy (e.g. 3CR~456
and 3CR~234).  As this method does not provide sound results for
objects with faint nuclei or at medium/high $z$, which represent the
majority of our sources, we developed and adopted a new procedure,
presented in the next section.

\subsection{Comparison with HST observations: 
  the color profile technique} \label{irlum}

The alternative approach we developed takes advantage of the fact that
HST images are available for the targets and thus the galaxy profile
and (when detected) the intensity of the optical nuclear source are
known a priori. Therefore, instead of modeling directly the IR images
we compared them to the HST ones (F702W R band filter).  The basic
idea is that an obscured nucleus would reveal itself through an
increase of the IR nuclear flux with respect to what is seen in the
optical images.

In order to perform the comparison it is necessary to produce a
synthetic HST image matching both the seeing and the pixel size of
each IR image. Therefore we fitted a Gaussian to all the stellar
objects in the HST field, measured their FWHM and adopted as FWHM the
median of all measurements with an uncertainty estimated from their
dispersion. The values obtained range between $0\farcs6$ and
$1\farcs7$ with a typical uncertainty of $\pm$7\% (the error on the
seeing is between $0\farcs02$ and $0\farcs2$).  The HST image has been
then convolved with a Gaussian with the appropriate FWHM and
interpolated to match the pixel size of NICS/ARNICA\footnote{We 
chose to use a Gaussian rather than a real PSF to convolve the HST 
images because the Gaussian model reproduced well the observed PSF. 
A small dependence of the PSF shape on the position in the FoV 
has been detected and accounted for in the error of the estimated Gaussian 
dispersion, hence allowing us to properly estimate the uncertainty on the 
derived nuclear NIR luminosity.}. 

The color-profile method is detailed in Appendix~\ref{colprofApdx} for
two show-case galaxies, namely 3CR~403 and 3CR~300. Briefly,
photometry is performed on annuli of increasing radius out to
3\arcsec~on the `matched' HST and on the reduced TNG images.  An
average color (and its associated error) is derived at each
annulus. The outer portion of each color profile sets the color of the
host galaxy.  The nuclear monochromatic $K^{\prime}$-band 
luminosity $L_{\rm K^{\prime}}$ is estimated as the sum of 
$L_{\rm K^{\prime},xs}$, i.e. the measured nuclear IR excess, and 
$L_{\rm K^{\prime},scaled}$, namely the HST optical nuclear component 
$L_{\rm O}$ properly scaled with the reference host galaxy color.

Also for all RGs (but two) with an optical nuclear upper limit, a
significant IR excess is present in the inner part of the color
profile.  Clearly in such cases $L_{\rm K^{\prime}}$ is less accurate:
$L_{\rm K^{\prime},scaled}$ is calculated considering $L_{\rm O}$ as a 
detection, and its error takes into account the possibility of $L_{\rm O}$ 
equal to zero. For 3CR~327 and 3CR~379.1 instead only an upper
limit to the nuclear IR excess could be estimated and therefore
$L_{\rm K^{\prime}}$ is also an upper limit. Four objects (namely
3CR~284, 3CR~321, 3CR~405 and 3CR~436) have optical complex nuclei. In
these cases, although no measurement is available for 
$L_{\rm K^{\prime},scaled}$ -- implying that the resulting 
$L_{\rm K^{\prime}}$ should be strictly considered as a lower limit -- the
`true' value would be very similar to the one estimated if 
$L_{\rm K^{\prime},scaled} \ll L_{\rm K^{\prime},xs}$. 

\subsubsection{Host galaxy properties} \label{host_prop}

\begin{figure*}
\centering
\includegraphics[width=18cm]{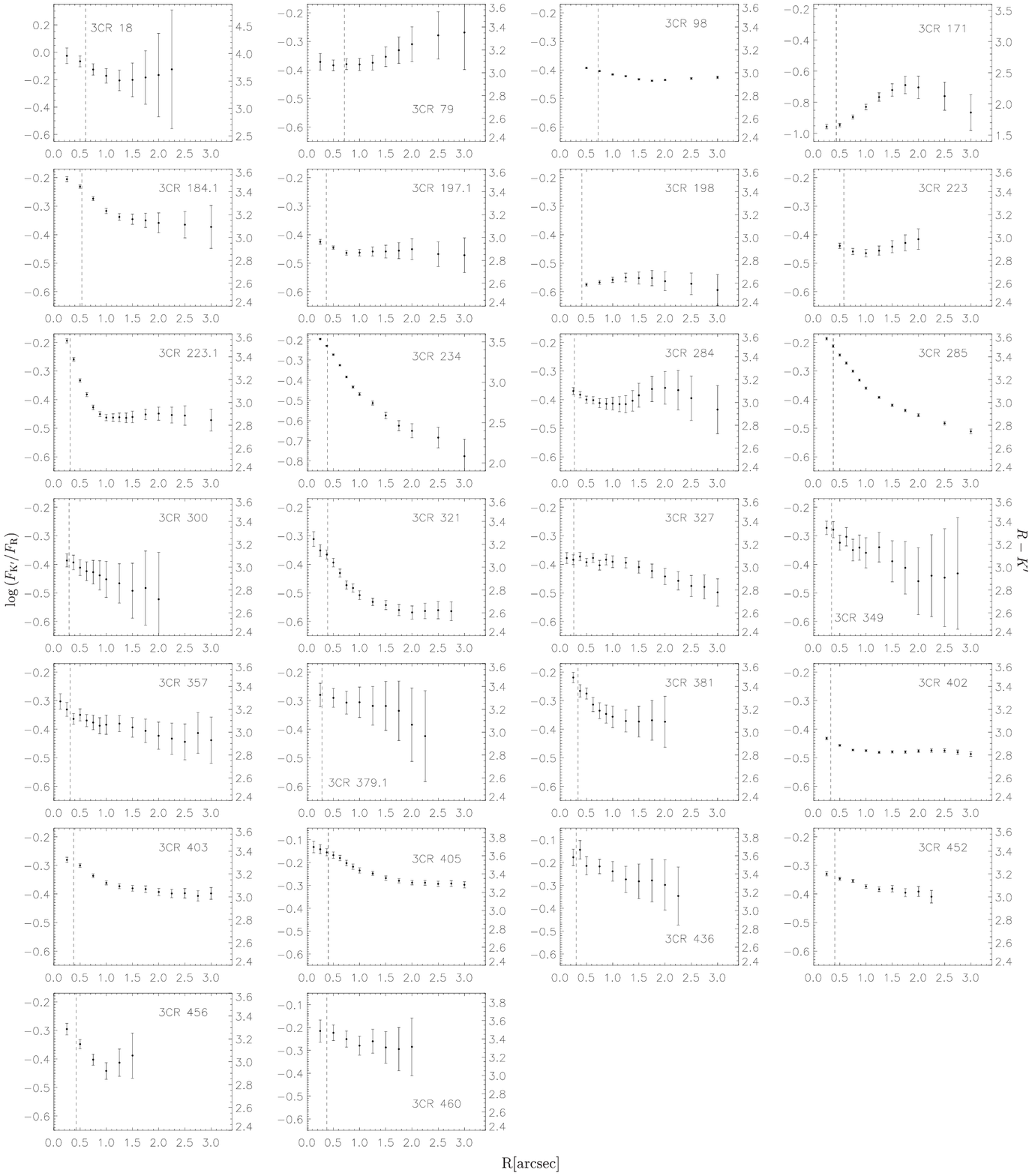}
\caption{Color radial profiles of the 26 HEG RGs [$R-K^{\prime}=
2.5\log{(F_{\rm K^{\prime}}/F_{\rm R})}+4.025$].  The error bars
represent the 1$\sigma$ error on the derived color, while the dashed
vertical line represents the dispersion (FWHM divided by 2.354) of the
PSF of the image.}
\label{profile}
\end{figure*}

In Fig.~\ref{profile} we present the color profile derived for each
source. For the majority of them the host galaxy color is consistent
with being constant; only a very few objects show significant
gradients or complex behaviour in the $R-K^{\prime}$ profile (see
Appendix~\ref{galnotes} for details on individual objects). For 19/26
the outer part of the profile has been fitted with a linear model
$\left.R-K^{\prime}\right|_{\rm model}=A+Br$ in order to account for a
possible gradient: 5/19 sources (3CR~98, 3CR~285, 3CR~349, 3CR~357 and
3CR~405) show significant (at $\ga$3$\sigma$) gradient, while other
7/19 objects (3CR~79, 3CR~223, 3CR~300, 3CR~379.1, 3CR~403, 3CR~436
and 3CR~452) have only marginally significant ($>$1$\sigma$) color
gradient.  The mean $R-K^{\prime}$ color of the host galaxy is
2.97$\pm$0.29, consistent with the value of $R-K$=2.75$\pm$0.45 of a
sample of RGs and RL QSOs hosts \citep{dunlop03} and with the expected
color $R-K$=2.84 for an elliptical at $z$=0.16 \citep{poggianti97}
(the mean redshift of the sample is 0.15). The mean value of the
$R-K^{\prime}$ gradient [defined as
$\Delta(R-K^{\prime})/\Delta(\log{r})$] is
-0.22$\pm$0.02~mag~dex$^{-1}$, consistent with the values of: i)
-0.14~mag~dex$^{-1}$ (with a broad distribution, from $-$0.62 to
$+$0.15~mag~dex$^{-1}$) of 12 nearby bright elliptical galaxies
(\citealt{peletier90a},b); ii) -0.35$^{+0.29}_{-0.42}$~mag~dex$^{-1}$
of a sample of early-type galaxies in the rich cluster AC~118 at
$z$=0.31 \citep{labarbera02}; iii)
$\Delta(R-H)/\Delta(\log{r})$=-0.2$\pm$0.1~mag~dex$^{-1}$ for the host
galaxies of BL~Lac objects, thought to be normal ellipticals
\citep{scarpa00} (the $H$ and $K$ host galaxy profiles should be
similar as dominated by slowly evolving red stars).  We conclude that
based on their colors and color gradients the hosts of the studied
sample of HEG RGs are perfectly consistent with being normal
ellipticals.

\subsubsection{IR nuclear measurements accuracy} \label{IRN_measure}

Clearly, effects other than nuclear obscuration might cause a color
change, such as extinction by extended dust structures or
age/metallicity gradients in the stellar population.  As stated above,
$\sim$50\% of the objects show gradients, with only $\sim$20\%
statistically significant. Such gradients indeed represent the main
limiting factor for the accuracy of the color profile method.
However, in most of the cases showing gradients, the estimates of
$L_{\rm K^{\prime}}$ based on the outer color gradient modeling or
assuming a flat profile differ by $\la$ 0.2~dex, and never more than a
factor $\sim$2 (see also Appendix~\ref{colprofApdx} for the case of
3CR~403).  Only 3CR~357 and 3CR~379.1 show larger differences, 0.44
and 0.57~dex, respectively; for the latter one, which has been
excluded from further analysis, only an upper limit for 
$L_{\rm K^{\prime}}$ could be estimated. Only 2 (out of 26) objects
(3CR~171\footnote{The source 3CR~171 is very peculiar, having a very
bright narrow-line emission region co-spatial to the radio emission
\citep{tadhunter00}, indicative of jet-cloud interaction; the
contribution to the HST image from line emission is large, if not
dominant \citep{dekoff00}. Its observed color profile (see
Fig.~\ref{profile}) is complex, with a central optical excess likely
due to the strong line emission contribution (see also
Appendix~\ref{galnotes} for details). We therefore neglect this object
in the following analysis.} and 3CR~284) show complex $R-K^{\prime}$
profiles. Therefore, we can conclude that for most sources the color
profile is sufficiently well behaved that our rather simple technique
provides robust estimates of the IR nuclear fluxes, with significantly
higher accuracy than the surface brightness decomposition method.

\begin{figure}
\resizebox{\hsize}{!}{\includegraphics{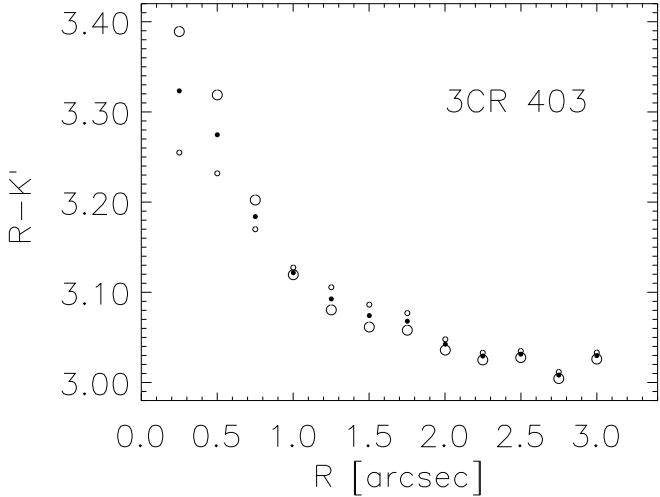}}
\caption{Color profiles of 3CR~403. The filled circles represent the
  profile obtained with the estimated FWHM of 0\farcs89; the large and
  small empty circles correspond to the profiles obtained varying the
  measured FWHM by 9\%, the variance of the different measurements of
  seeing. The difference is $<$ 7\% in the inner most point, $<$4\% at
  a radius of 0\farcs5 and completely negligible at $>$ 1\arcsec.}
\label{test}
\end{figure}

Before proceeding any further, it is also crucial to assess the effect
of the uncertainty in the seeing measurement on the color profile as
this is the only free parameter of the analysis. In Fig.~\ref{test} we
compare profiles obtained varying the measured FWHM of 3CR~403 by 9\%
(corresponding to the variance of the different measurements of seeing
for this source, slightly larger than that typical of the studied
sample).  The differences between the profiles are only marginal,
being less than $\pm$ 0.04 in color at 0\farcs5 and completely
negligible at 1\arcsec. As detailed in Appendix~\ref{colprofApdx},
this implies an uncertainty on the nuclear IR luminosity in the range
0.02-0.2~dex depending on the specific source.

A crucial test for the robustness of our procedure is given by the
comparison with the results from direct imaging for the three sources
for which HST-NICMOS observations are publicly available, namely
3CR~405, 3CR~456 and 3CR~460 (for details see Appendix~\ref{galnotes}).
For 3CR~405, \citet{tadhunter99} found a nuclear point-like source with 
a luminosity of (4.1$\pm$0.2)$\times$10$^{41}$ \ergs~at
2.25 $\mu$m.  In the hypothesis of a flat outer $R-K^{\prime}$
profile, we estimated a nuclear luminosity of
(7.3$\pm$0.9)$\times$10$^{41}$ \ergs~at 2.12 $\mu$m, about 0.25~dex
larger.  However, by modeling the small color gradient at radii
$>$1\farcs25, we obtained 
$\nu L_{\rm K^{\prime}} \sim$10$^{41.66\pm0.09}$ \ergs~, perfectly 
consistent with the NICMOS result. For 3CR~456 and 3CR~460 NICMOS-NIC2 
images are available in the public archive at STScI. A central point-like 
source is clearly present in 3CR~456 with a luminosity of 
10$^{43.63\pm0.09}$ \ergs~at 2.05 $\mu$m. After modeling the outer color
gradient, we estimated $\nu L_{\rm K^{\prime}}=$ 10$^{43.46\pm0.04}$
\ergs~, within $\sim$2$\sigma$ from the HST value.  In the
HST image of 3CR~460 only extended emission is visible. An upper limit
on the IR nuclear luminosity of 10$^{42.82}$ \ergs ~at 2.05 $\mu$m has
been estimated, consistent with our measurement of 
$\nu L_{\rm K^{\prime}}$=10$^{42.79\pm0.19}$ \ergs.  Considering the
different methods, aperture and spatial resolution of the above
observations, the agreement between the HST images and the color
profile analysis results is quite remarkable. Early results from
new HST/NICMOS observations confirm our measurements for 3CR~79,
3CR~379.1 and 3CR~403 (M. Chiaberge, priv. comm).

It thus turns out that the $R-K^{\prime}$ color radial profile method
returns values of $L_{\rm K^{\prime}}$ correct within at most a factor
of 2, and most likely with an average error of 0.2~dex in log, even
for very faint nuclei (e.g. 3CR~405). The high accuracy basically
follows from the information provided by the high spatial resolution
optical HST imaging. In Table~\ref{tab2} the measured values of the
nuclear $\nu L_{\rm K^{\prime}}$ for the HEG sample are listed; the
details of the analysis for the individual objects can be found in
Appendix~\ref{galnotes}.

\begin{table*}
\caption{Nuclear luminosities and derived quantities}
\small
\begin{tabular}{l c c c c} \hline
Name &   $\log{(\nu L_{\rm K^{\prime}})}$$^{a}$ & $A_{\rm V}$$^{b}$ & $f$$^{c}$ &  $A_{\rm V,torus}$$^{d}$ \\
     &             \ergs                        &          (mag)    &     (\%)  & (mag) 		\\
 (1) &  (2)                                     &  (3)              &  (4)      &  (5)  		\\
     &                                          &                   &           &       \\
\hline                                		  
3C~18    &   43.66$\pm$0.06           &    1.87$\pm$0.27      & $\sim$0                & 1.9$\pm$0.3           \\
3C~33    &        --                  &   	--		   & --                     &    --                 \\
3C~63    &        --                  &   	--		   & --       	            &    --                 \\
3C~79    &   42.87$\pm$0.21           &   -0.12$\pm$0.87 	   & 5.5$^{+3.5}_{-3.4}$    & 60$^{+\infty}_{-29}$  \\
3C~98    &   41.80$^{+0.03}_{-0.09}$  & $>$6.19$\pm$0.28 	   & $<$0.43                & 17.0$^{+1.8}_{-1.3}$  \\
3C~105   &        --                  &   	--		   & --       	            &    --                 \\
3C~135   &        --                  &   	--		   & --       	            &    --                 \\
3C~153   &        --                  &   	--		   & --       	            &    --                 \\
3C~171   &        --                  &   	--		   & --       	            &    --                 \\
3C~184.1 &   43.16$\pm$0.04       	   &    1.67$\pm$0.30      & 7.0$^{+2.3}_{-1.7}$    & 19.5$^{+2.0}_{-1.7}$  \\
3C~192   &        --                  &   	--		   & --       	            &    --                 \\
3C~197.1 &   42.81$\pm$0.07           &    0.56$\pm$0.38  	   & $\sim$0                & 0.6$\pm$0.4	    \\
3C~198   &   42.57$\pm$0.10           &   -0.12$\pm$0.47  	   & $\sim$0                & -0.1$\pm$0.5          \\
3C~223   &   42.43$^{+0.05}_{-0.15}$  & $>$2.34$\pm$0.33      & $<$0.88                & 33.2$^{+3.0}_{-2.7}$  \\
3C~223.1 &   43.03$\pm$0.04    	      & $>$5.08$\pm$0.30      & $\sim$0                & $>$5.1$\pm$0.3        \\
3C~234   &   44.06$\pm$0.04           &    1.55$\pm$0.30      & 6.3$^{+2.0}_{-1.6}$    & 21.5$^{+2.2}_{-1.8}$  \\
3C~284   &   $>$42.34$\pm$0.33        &   	--		   & --                     &    --                 \\
3C~285   &   42.35$\pm$0.04           &    8.46$\pm$0.30 	   & $\sim$0                & 8.5$\pm$0.3           \\
3C~300   &   42.54$\pm$0.16           &    0.71$\pm$0.69  	   & 3.0$^{+3.7}_{-1.6}$    & 37.7$^{+32.3}_{-7.7}$ \\
3C~321   &   $>$42.73$\pm$0.06        &   	--		   & --                     &    --                 \\
3C~327   &   $<$41.77                 & $\leogr$2.58$\pm$0.29 & $<$0.22                & 44$^{+\infty}_{-2}$   \\
3C~349   &   43.02$\pm$0.11           &    1.15$\pm$0.50      & $\sim$0                & 1.2$\pm$0.5           \\
3C~357   &   42.19$\pm$0.35           & $>$2.86$\pm$1.41      & $<$1.0                 & 29$^{+10}_{-9}$     \\
3C~379.1 &   $<$42.04                 & $\leogr$1.03$\pm$0.28 & --                     &    --                 \\
3C~381   &   42.98$^{+0.07}_{-0.10}$  & $>$3.53$\pm$0.38      & $<$1.0                 & 24.5$^{+2.0}_{-1.2}$  \\
3C~402   &   41.59$\pm$0.06           &    1.71$\pm$0.35      & --                     &    --                 \\
3C~403   &   42.27$\pm$0.05           &    4.17$\pm$0.33      & 0.88$^{+0.32}_{-0.23}$ & 21.9$\pm$1.1          \\
3C~405   &   $>$41.66$\pm$0.09        &   	--		   & --                     &    --                 \\
3C~436   &   $>$42.53$\pm$0.36        &   	--		   & --                     &    --                 \\
3C~452   &   42.04$^{+0.10}_{-0.23}$  & $>$2.30$\pm$0.47      & $<$2.5                 & 24.0$^{+6.0}_{-4.5}$  \\
3C~456   &   43.46$\pm$0.04           &    1.55$\pm$0.30 	   & 6.0$^{+2.0}_{-1.3}$    & 22.0$^{+2.0}_{-1.7}$  \\
3C~460   &   42.79$\pm$0.19           &    2.46$\pm$0.80	   & 6.3$^{+10.2}_{-3.8}$   & 14.7$^{+5.8}_{-5.2}$  \\
\hline
\end{tabular}

{
$^{a}$ $\log{(\nu L_{\rm K^{\prime}})}$: total luminosity at 2.12~$\mu$m 
estimated as in \S~\ref{irlum};

$^{b}$ Nuclear extinction from eq.~\ref{eq_AV};

$^{c}$ Fraction (percentage) of the total nuclear emission scattered 
toward the line-of-sight (from Fig.~\ref{abs_scatt2}); $f \sim$0 
means that the emission is consistent with transmitted light both in 
optical and NIR;

$^{d}$ Torus extinction (from Fig.~\ref{abs_scatt2}).
}

\label{tab2}

\end{table*}




\section{Results}
\label{results}

\subsection{Nuclear properties of narrow and broad lined RGs}

\begin{figure}
\resizebox{\hsize}{!}{\includegraphics{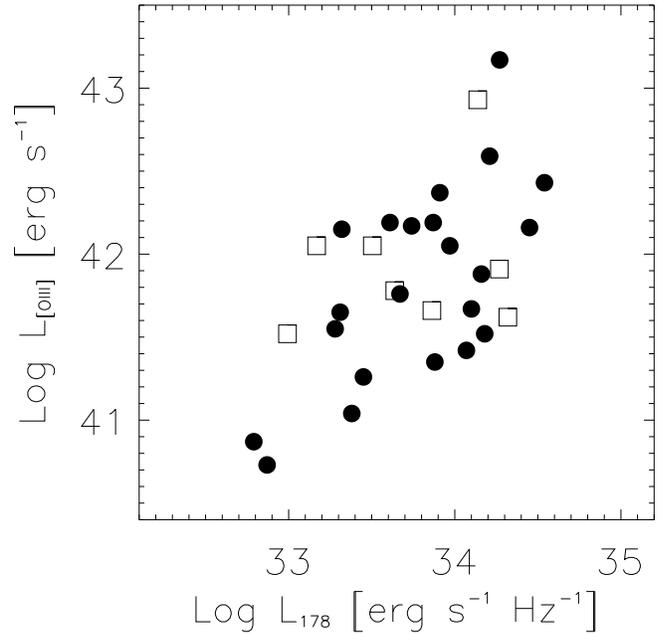}}
\caption{[OIII] line luminosity, $L_{\rm [OIII]}$, versus total radio
  luminosity at 178~MHz, $L_{\rm 178}$. HEGs are represented by filled
  circles and BLRGs by empty squares.}
\label{L178_Lr_LOIII_heg_blrg}
\end{figure}

As pointed out in \S~1, in the standard unification picture, HEGs are
believed to be intrinsically identical to BLRGs, and obscuration by a
dusty torus is thought to be responsible for their different nuclear
properties.  In this scenario HEGs should be indistinguishable from
BLRGs when comparing isotropic quantities such as the [OIII] line
luminosity, $L_{\rm [OIII]}$, and the total radio luminosity at
178~MHz, $L_{\rm 178}$. This indeed appears to be the case:
Fig.~\ref{L178_Lr_LOIII_heg_blrg} shows $L_{\rm 178}$ versus 
$L_{\rm [OIII]}$ for the complete samples of HEGs and BLRGs drawn from the
3CR sample within $z \la$0.3 (from CCC02).  The Kolmogorov-Smirnoff 
test and the mean values of the two quantities are consistent with the
two populations being drawn from the same parent distribution (see
Tab.~\ref{tab_heg_blrg} for a summary of the statistical comparisons
between HEGs and BLRGs).

\begin{table}
\caption{Results of statistical tests: comparison of means and 
standard deviations of different variables for HEGs and
BLRGs; $p$ is the probability that the two samples are drawn 
from the same parent distribution according to the Kolmogorov-Smirnoff test.}
\small
\begin{tabular}{c c c c} 
\hline
 \multicolumn{4}{c}{} \\
 Sample & HEG & BLRG &  $p$ \\
\hline
$\log{L_{\rm 178}}$   & 33.77$\pm$0.48 & 33.74$\pm$0.50 & 0.99  \\
$\log{L_{\rm [OIII]}}$& 41.83$\pm$0.59 & 41.94$\pm$0.45 & 0.65  \\
$\log{L_{\rm r}}$     & 30.77$\pm$0.75 & 31.37$\pm$0.42 & 0.06  \\
$\log{L_{\rm O}}$  & 27.79$\pm$0.96* & 28.92$\pm$0.57 & 6$\times$10$^{-4}$* \\
$\log{L_{\rm K^{\prime}}}$  & 28.57$\pm$0.63* & 29.34$\pm$0.57 & 0.001*  \\
$\log{L_{\rm O,~A_V~corr}}$ & 28.38$\pm$0.63* &  ---  & 0.04*    \\
$\log{L_{\rm K^{\prime},~A_V~corr}}$ & 28.80$\pm$0.63* &  ---  & 0.04*  \\
$\log{\nu L_{\rm 60}}$      & 44.8$\pm$0.4  & 44.3$\pm$0.4  & 0.13-0.18 \\
\hline
\end{tabular}

{* only detections for HEGs are considered; in this case, the 
probabilities that the observed $\log{L_{\rm 178}}$ and $\log{L_{\rm [OIII]}}$ 
distributions are drawn from the same parent population are $p$=0.89 
and $p$=0.57, respectively.}

\label{tab_heg_blrg}
\end{table}

\begin{figure*}
\centering \includegraphics[width=18cm]{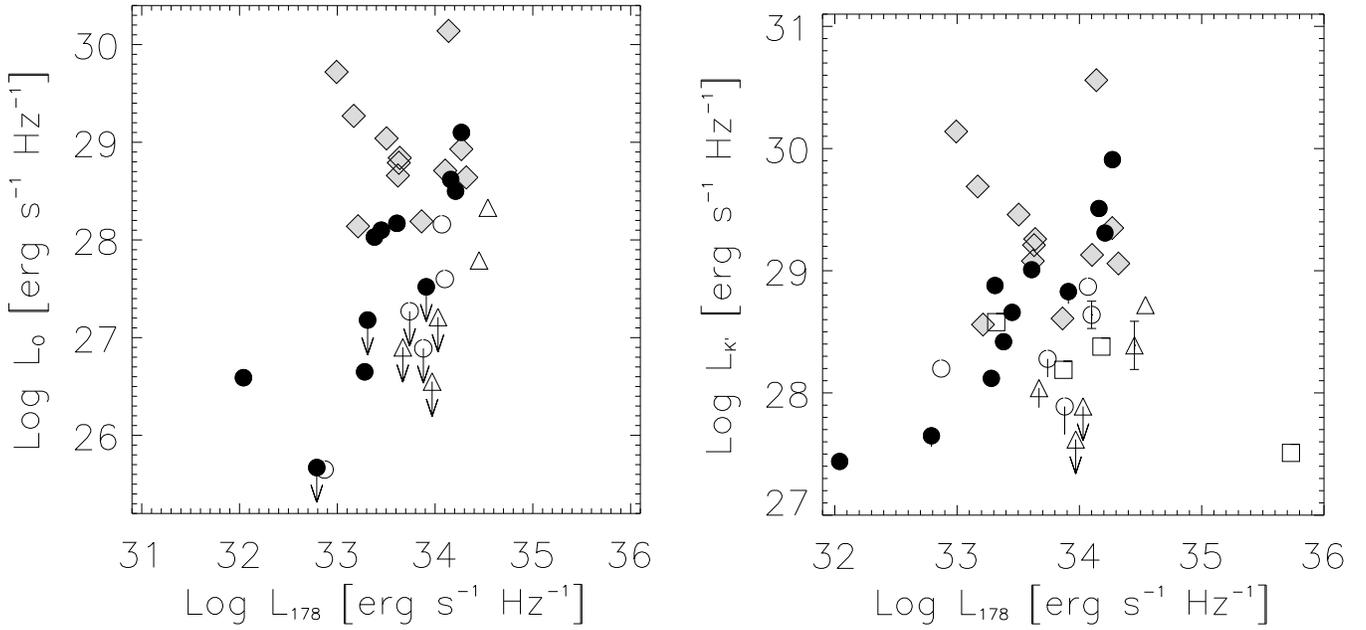}
\caption{{\bf Left panel:} Optical core luminosity $L_{\rm O}$ vs
$L_{\rm 178}$. {\bf Right panel:} NIR nuclear luminosity $L_{\rm
K^{\prime}}$ vs $L_{\rm 178}$. BLRGs are represented by light filled
diamonds; the different symbols to represent HEGs correspond to the
different quality of the $R-K^{\prime}$ color profile: filled circles
for good quality, triangles for worst quality and empty circles for
intermediate one. Typical error bars for the three quality levels are
also shown; arrows are upper limits and the short vertical lines
associated to the few objects in the right-hand panel indicate the
value of $L_{\rm K^{\prime}}$ if their $L_{\rm O}$ where equal to zero
(since the latter are only upper limits). Squares represent HEGs
having complex optical morphologies (no upper limits on the optical
core luminosities could be estimated).}
\label{LoLk_L178}
\end{figure*}

This unification scenario can now be further tested by including the
information on the nuclear NIR luminosities.  In Fig.~\ref{LoLk_L178}
we have plotted the nuclear optical and IR luminosity, $L_{\rm O}$ and
$L_{\rm K^{\prime}}$ versus $L_{\rm 178}$, for both HEGs and
BLRGs\footnote{We have chosen to use the isotropic $L_{\rm 178}$
instead of $L_{\rm r}$ (as done by CCC02) since the latter is likely
affected by beaming (BLRGs show brighter radio cores with respect to
HEGs -- see Tab.~\ref{tab_heg_blrg}).}.  As already shown by CCC02,
BLRGs have systematically larger $L_{\rm O}$ with respect to HEGs:
after excluding the upper limits (i.e.  optically undetected HEGs)
BLRGs are brighter on average by a factor $\sim$13 (including HEG
upper limits increases the difference). Essentially the same result
holds for the IR nuclear luminosity (see Fig.~\ref{LoLk_L178}, right
panel): the BLRG sub-sample is characterized by larger $L_{\rm
K^{\prime}}$\footnote{For BLRGs $L_{\rm K^{\prime}}$ has been
estimated from $L_{\rm O}$ (from CCC02) assuming an
optical-to-infrared spectral index typical of QSOs \citep{elvis94}.}
with respect to HEGs even though by a factor ($\ga$ 6) smaller than in
the optical band.

In order to assess whether and to which degree the different
optical/IR luminosities could be ascribed to absorption, we estimate
the amount of obscuration ($A_{\rm V}$) from the comparison of the
observed ratio $\left. L_{\rm O}/L_{\rm K^{\prime}}\right|_{\rm obs}$
with an intrinsic (assumed) value of such ratio, $\left. L_{\rm
O}/L_{\rm K^{\prime}}\right|_{\rm int}$. More specifically:
\begin{equation} 
\label{eq_AV}
A_{\rm V} = \frac{1}{0.252} \Bigg[ \left.\log{\frac{L_{\rm O}}
{L_{\rm K^{\prime}}}}\right|_{\rm int} - 
\left.\log{\frac{L_{\rm O}}{L_{\rm K^{\prime}}}}\right|_{\rm obs} \Bigg],
\end{equation}
where $\left.\log{L_{\rm O}/L_{\rm K^{\prime}}}\right|_{\rm int}
\sim$-0.42\footnote{With a typical error of $\sim$0.2~dex
\citep{elvis94}.} is assumed equal to that typical for RL QSOs
\citep{elvis94} (and corresponds to an optical-to-infrared spectral
index of $\sim$0.87).  The estimated extinctions are listed in
Table~\ref{tab2} (column 3): their range is $0 < A_{\rm V} < 4$ mag,
with only 3 exceptions.

From these values of $A_{\rm V}$ an extinction corrected nuclear
optical luminosity $L_{\rm O,~A_V~corr}$ can be inferred from 
$A_{\rm R}=0.751\, A_{\rm V}$ \citep{cardelli89}. 
Even after correcting for the estimated nuclear extinction, the BLRG
optical luminosities systematically exceed those of HEGs (see
Tab.~\ref{tab_heg_blrg}).  Clearly, this still holds for the NIR
nuclear luminosity, only marginally affected by the relatively small
$A_{\rm V}$ (see Tab.~\ref{tab_heg_blrg}).
Thus, based on the optical/NIR nuclear continuum properties, HEGs and
BLRGs do not seem to belong to the same parent population even after
absorption (inferred from the ratio between optical and NIR nuclear
emission) is taken into account.  This seems to imply that HEGs could
be intrinsically fainter than their putative BLRGs counterparts
(having similar narrow line and extended radio luminosities).

\subsection{Nuclear and emission line properties}

This result appears however to be in contradiction with the similarity
in the $L_{\rm [OIII]}$ distributions for HEGs and BLRGs, as the same
intrinsic nuclear radiation field would be expected to maintain
similar gas photo-ionization and line luminosities (if indeed
photoionization by the nuclear radiation field is indeed the chief
mechanism producing $L_{\rm [OIII]}$).

\begin{figure}
\resizebox{\hsize}{!}{\includegraphics{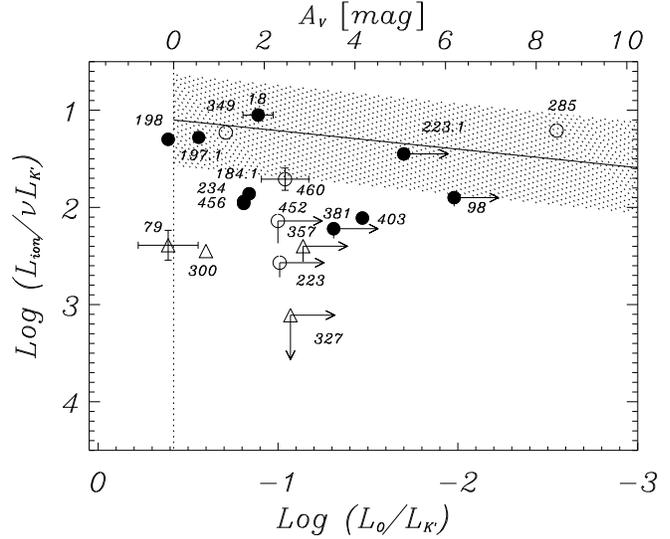}}
\caption{Ratio between the ionization luminosity $L_{\rm ion}$ and 
$\nu L_{\rm K^{\prime}}$ versus the ratio between $L_{\rm O}$ and
$L_{\rm K^{\prime}}$.  Symbols as in Fig.~\ref{LoLk_L178}.  The arrows
represent upper/lower limits. The dotted vertical line indicates the
typical QSO ratio for $\log{(L_{\rm O}/L_{\rm K^{\prime}})}$
\citep{elvis94}. The continuous line represents the value 
$L_{\rm ion}/\nu L_{\rm K^{\prime}}$ (estimated for BLRGs) when absorption 
is taken into account [i.e. 
$\log{(L_{\rm ion}/\nu L_{\rm K^{\prime}})} \sim 1.1 + 0.048 A_{\rm V}$, 
$A_{\rm V}$ from eq.~\ref{eq_AV}] and the shaded area shows the location 
of BLRGs (estimated from the dispersion of $\sim$0.47~dex around the mean 
value).  The upper axis, labelled $A_{\rm V}$, is obtained from the lower 
one via eq.~\ref{eq_AV}.}
\label{abs_scatt}
\end{figure}

In such hypothesis, for more quantitative estimates we report in
Fig.~\ref{abs_scatt} the ratio between $\nu L_{\rm K^{\prime}}$ (from
Table~\ref{tab2}) and the ionizing luminosity $L_{\rm ion}$.  $L_{\rm
ion}$ has been estimated from the total luminosity in narrow lines
$L_{\rm NLR}$ [inferred from $L_{\rm [OIII]}$ (Table~\ref{tab1}) as
$L_{\rm NLR}=3(3L_{\rm [OII]}+1.5L_{\rm [OIII]})$ and $L_{\rm
[OIII]}=4L_{\rm [OII]}$; e.g.  \citealt{rawlings91}], as $L_{\rm
ion}=L_{\rm NLR} C^{-1}$, adopting a covering factor $C
\sim$0.01\footnote{Note that the precise value of $C$ is ininfluent in
the following analysis.}.  For BLRGs, $\log{(L_{\rm ion}/\nu L_{\rm
K^{\prime}})} \sim$1.1$\pm$0.47.

When compared with the typical values for BLRGs, $L_{\rm ion}/\nu
L_{\rm K^{\prime}}$ thus represents the `discrepancy' between the
nuclear emission as inferred from the lines and the estimated $L_{\rm
K^{\prime}}$ for HEGs. The representative values for BLRGs are
indicated in Fig.~5 by the shaded area (accounting for the dispersion
of $\sim$0.47~dex of BLRGs around their mean value shown as continuous
line).  As $L_{\rm ion}/\nu L_{\rm K^{\prime}}$ is affected by
obscuration [as $\log{(L_{\rm ion}/\nu L_{\rm K^{\prime}})} \sim 1.1 +
0.048 A_{\rm V}$], we plotted this ratio against the observed $L_{\rm
O}/L_{\rm K^{\prime}}$, which corresponds to $A_{\rm V}$ (reported on
the upper axis).  Obscuration then `moves' the position of sources in
this plane along a line parallel to the continuous one.  Therefore, if
nuclear obscuration is the only cause of the differences between HEGs
and BLRGs, then HEGs should be located within the shaded area,
depending on the amount of obscuration reflected in their optical-NIR
colors.  However, only $\sim$30\% of HEGs are consistent with this
picture.  For most sources, instead, the observed $L_{\rm ion}/\nu
L_{\rm K^{\prime}}$ is much larger than that typical of BLRGs -- even
when extinction is considered -- up to factors larger than 20 (for
detections).  We conclude that obscuration alone cannot account for
the differences in the optical/NIR nuclear properties of HEGs and
BLRGs with respect to the similarity in their narrow line emission.

\subsection{The role of scattering in HEGs}

An ingredient we have so far neglected is the possible presence of a
nuclear component due to scattered light.  As reported by CCC02, there
are at least two sources (3CR~234 and 3CR~109) for which
spectropolarimetric studies indicate that their optical nuclear
component corresponds to a compact scattering region, whose luminosity
indeed matches the optical nuclear emission seen in the HST images.
CCC02 argued that the HST optical cores in most HEGs can indeed be
produced by scattering.

The presence of a scattered component radically changes the
interpretation of our results. Let us examine the consequences of this
hypothesis in a simple scenario, where the observed nuclear luminosity
is simply due to direct light attenuated by a dusty torus
characterized by an extinction $A_{\rm K^{\prime}, torus}$, and a fraction
$f$ of the total nuclear light ($\left.L_{\rm K^{\prime}}\right|_{\rm int}$)
scattered toward the line-of-sight. E.g. in the $K^{\prime}$ band:
\begin{equation} \label{eq_nukeIR_2comp}
\left.L_{\rm K^{\prime}}\right|_{\rm obs} = \left.L_{\rm K^{\prime}}\right|_{\rm int}
10^{-0.4 \cdot A_{\rm K^{\prime}, torus}} + f \left.L_{\rm K^{\prime}}\right|_{\rm int}.
\end{equation}

For $A_{\rm R}$=0.751$A_{\rm V}$ and $A_{\rm K^{\prime}}$=0.12$A_{\rm V}$
then:

\begin{equation} \label{eq_final1}
\left.\frac{L_{\rm O}}{L_{\rm K^{\prime}}}\right|_{\rm obs} =
\left.\frac{L_{\rm O}}{L_{\rm K^{\prime}}}\right|_{\rm int} \Bigg(
\frac{10^{-0.3 \cdot A_{\rm V, torus}} + f}
{10^{-0.048 \cdot A_{\rm V, torus}} + f}   \Bigg),
\end{equation}
and
\begin{equation} \label{eq_final2}
\left.\frac{L_{\rm ion}}{\nu L_{\rm K^{\prime}}}\right|_{\rm obs} =
\left.\frac{L_{\rm ion}}{\nu L_{\rm K^{\prime}}}\right|_{\rm int} 
\Bigg({\frac{1 + f}
{10^{-0.048 \cdot A_{\rm V, torus}} + f}} \Bigg),
\end{equation}
where $\left.\log{(L_{\rm O}/L_{\rm K^{\prime}})}\right|_{\rm int} \simeq$ -0.42 and 
$\left.\log{(L_{\rm ion}/\nu L_{\rm K^{\prime}})}\right|_{\rm int} \simeq$ 1.1$\pm$0.47 
are the typical values for BLRGs.

\begin{figure*}
\centering
\resizebox{\hsize}{!}{\includegraphics{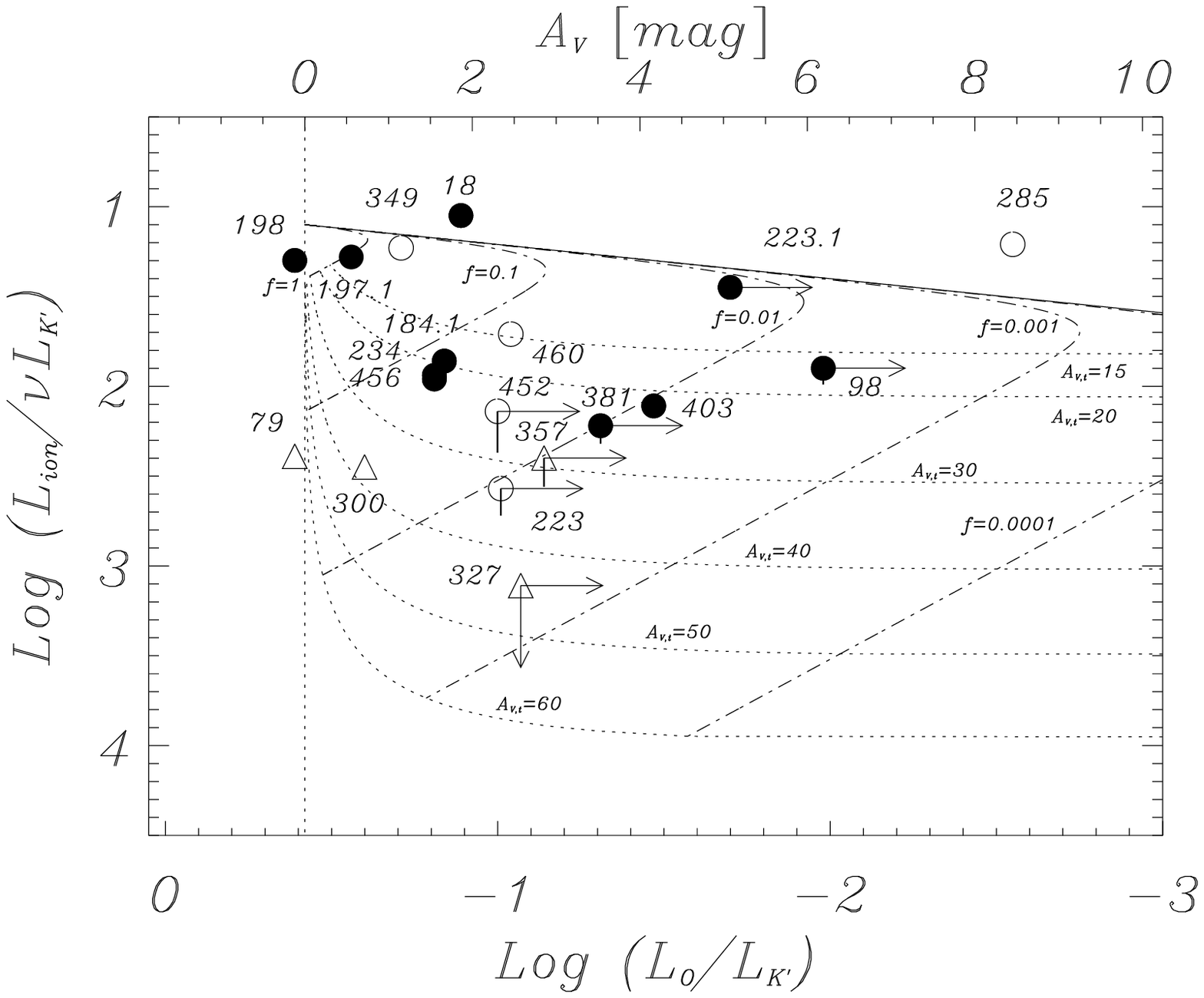}}
\caption{Same as Fig.~\ref{abs_scatt}, with over plotted the `lines'
  $\left.L_{\rm ion}/\nu L_{\rm K^{\prime}}\right|_{\rm obs}$ as
  function of $\left.L_{\rm O}/L_{\rm K^{\prime}}\right|_{\rm obs}$,
  obtained from eq.~\ref{eq_final1} and ~\ref{eq_final2}, for several
  values of $f$ and $A_{\rm V,~torus}$.  The dot-dashed lines
  correspond to fixed value of $f$ and running $A_{\rm V,~torus}$, and
  vice-versa for the dotted lines (as labelled).}
\label{abs_scatt2}
\end{figure*}

The effect of the two added (free) parameters, $A_{\rm V, torus}$ and 
$f$, on $L_{\rm ion}/\nu L_{\rm K^{\prime}}$ and 
$L_{\rm O}/L_{\rm K^{\prime}}$ -- i.e. the plane defined in 
Fig.~\ref{abs_scatt} -- can be graphically 
seen in Fig.~\ref{abs_scatt2} as dotted and dot-dashed lines (obtained
from eq.~\ref{eq_final1} and ~\ref{eq_final2} by fixing $A_{\rm V, torus}$ 
and letting $f$ run, and vice-versa). As an illustrative
example, let us concentrate on the line representing $f=0.01$.  For
low values of obscuration ($A_{\rm V,~torus} < 5$ mag) the transmitted
light exceeds the scattered fraction both in the optical and NIR. A
source `moves' along the solid tilted line simply describing the
effects of increasing nuclear obscuration. For larger values of
$A_{\rm V,~torus}$ the transmitted optical component becomes
increasingly negligible with respect to the scattered one and $L_{\rm O}$ 
is essentially constant and equal to $f \left.L_{\rm O}\right|_{\rm int}$. 
On the other hand $L_{\rm K^{\prime}}$ keeps
decreasing, i.e. $L_{\rm ion}/\nu L_{\rm K^{\prime}}$ increases, causing the
turnover toward the bottom left corner of the diagram. Only for values
of $A_{\rm V,~torus} > 40$ mag the scattered and transmitted $K^{\prime}$-band
light become comparable: the `track' then halts to the intrinsic
$L_{\rm O}/L_{\rm K^{\prime}}$ ratio with luminosities reduced to a fraction
$f$ on the intrinsic ones.

The properties of all HEGs reported in Fig.~\ref{abs_scatt2} can then
be interpreted with different contributions of transmitted and
scattered light: their location in Fig.~\ref{abs_scatt2} translates
into constraints on $f$ and $A_{\rm V,torus}$ (listed in
Table~\ref{tab2} for each source).  Only 6 sources, located close to
the continuous line, are consistent with being BLRGs seen through a
small amount of dust, with $A_{\rm V,torus} < 8.5$ mag.  The majority of
HEGs require larger obscuration, namely 15 $< A_{\rm V,torus} <$ 50
mag, and their optical cores can only be due to scattering.  The
scattered fraction ranges between $\sim$ 0.2 and 7\%: indeed HEGs with
detected optical nucleus in the HST images have $f > 0.9\%$, while
objects for which no optical HST source is seen correspond to upper limits
in the range 0.2 - 2.5\%.  Clearly the values of extinction derived from
the $L_{\rm O}/L_{\rm K^{\prime}}$ ratio (\S~4.1) systematically
underestimate the nuclear absorption.

Summarizing: within this simple scenario, for the majority 
($\sim$70\%) of HEGs both nuclear
absorption by a dusty torus and scattering of the nuclear light are
required to unify them with BLRGs. At optical wavelengths, the
scattered component usually dominates (except for the few sources
which are characterized by a low $A_{\rm V,torus}$), while in the NIR
the nuclear light is generally transmitted through the torus,
attenuated by a factor typically $\sim$10 but up to $\sim 100$.

So far we have neglected the effects of redshift, having assumed that
all HEGs are at $z$=0 in order to avoid introducing uncertainties due
to the unknown spectral shape. In order to quantify these effects on
our results, we have repeated the above analysis assuming $z$=0.15,
which is the median redshift of the BLRG and HEG populations (the $z$
distribution of the two population are statistical identical, with a
Kolmogorov-Smirnov probability $p$=0.81). At $z$=0.15, the observed
optical and NIR luminosities correspond to rest-frame wavelengths 0.6
and 1.85 $\mu$m, respectively. Consequently, we have re-estimated the
rest-frame values for $\left.L_{\rm O}/L_{\rm K^{\prime}}\right|_{\rm
int}$ and $\left.L_{\rm ion}/ \nu L_{\rm K^{\prime}}\right|_{\rm
int}$, and the extinction at the appropriate wavelengths. The newly
estimated values for $A_{\rm V,torus}$ are smaller by $\sim$25\%
compared to those for $z$=0 (the rest-frame bands correspond to
shorter wavelengths, where the extinction is higher), the fractions
$f$ are almost unaffected. No trend of $A_{\rm V,torus}$ with $z$ is
present. We conclude that our results are not significantly affected
by the $z$=0 assumption adopted.

\begin{figure}
  \resizebox{\hsize}{!}{\includegraphics{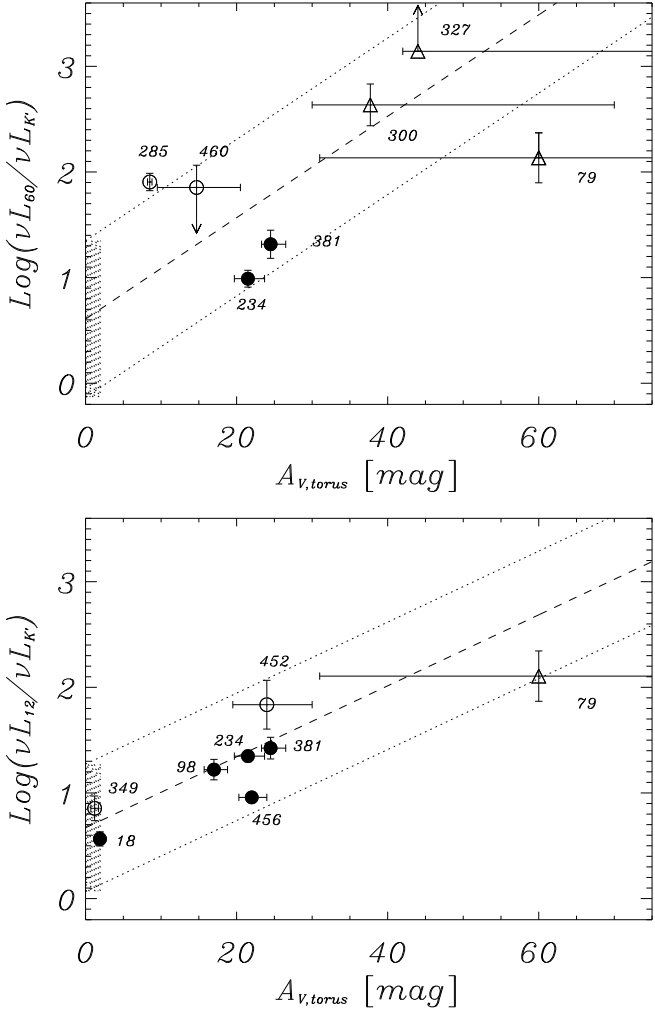}}
\caption{{\bf Upper panel:} Ratio between the 60~$\mu$m FIR luminosity
$\nu L_{\rm 60}$ and $\nu L_{\rm K^{\prime}}$ (2.12 $\mu$m) versus the
torus extinction $A_{\rm V,torus}$ (from Table~\ref{tab2}). {\bf Lower
panel:} Ratio between the 12~$\mu$m MIR luminosity $\nu L_{\rm 12}$
and $\nu L_{\rm K^{\prime}}$ versus $A_{\rm V,torus}$. Symbols as in
Fig.~\ref{abs_scatt}. The shaded area is the BLRG locus: the mean
values are $\nu L_{\rm 60}/\nu L_{\rm K^{\prime}}
\sim$0.61$\pm$0.74~dex and $\nu L_{\rm 12}/\nu L_{\rm K^{\prime}}
\sim$0.67$\pm$0.60~dex. The dashed lines in both panels represent the
BLRG value when absorption in the NIR is taken into account
[$\log{(\nu L_{\rm 60}/\nu L_{\rm K^{\prime}})} \sim 0.61 + 0.048
A_{\rm V,torus}$] (upper panel) and when absorptions both in the NIR
and MIR are taken into account [$\log{(\nu L_{\rm 12}/\nu L_{\rm
K^{\prime}})} \sim 0.67 + 0.034 A_{\rm V,torus}$] (lower panel); the
two dotted lines delimit the BLRG `region'.}
\label{fig-FIR}
\end{figure}

An important test of our findings is provided by FIR information, as
the FIR luminosity should be orientation-independent. A few HEGs and
BLRGs in our sample have published FIR (60~$\mu$m) luminosities ($\nu
L_{\rm 60}$) (\citealt{heckman94}; \citealt{haas04}). Within the large
uncertainties and low statistics, the distributions of $\nu L_{\rm
60}$ for HEGs and BLRGs are consistent with being drawn from the same
population (see Table~\ref{tab_heg_blrg}) and the observed $\nu L_{\rm
60}/\nu L_{\rm K^{\prime}}$ ratio of HEGs is consistent with that of
BLRGs once the extinction $A_{\rm V,torus}$ due to the dusty torus is
taken into account, as shown in Fig.~\ref{fig-FIR} (upper panel).
The same test can be performed using MIR (12~$\mu$m) luminosities
($\nu L_{\rm 12}$) (\citealt{heckman94}; \citealt{siebenmorgen04}). In
Fig.~\ref{fig-FIR} (lower panel) the observed ratio $\nu L_{\rm
12}/\nu L_{\rm K^{\prime}}$ is plotted instead of $\nu L_{\rm 60}/\nu
L_{\rm K^{\prime}}$. Absorption in the MIR has been taken into account
by using the extinction law with standard graphite-silicate mixes
(e.g. \citealt{draine89}), which implies $A_{\rm 12}=0.036A_{\rm
V}$. Once the extinction $A_{\rm V,torus}$ is taken into account, we
find full consistency between HEGs and BLRGs (the probability that the
$\nu L_{\rm 12}$ distributions of HEGs and BLRGs are drawn from the
same parent population is $p$=0.47, when luminosities are corrected
for extinction, with $<\nu L_{\rm 12}>$=10$^{44.23\pm0.23}$ for BLRGs
and $<\nu L_{\rm 12}>$=10$^{44.59\pm0.88}$ for HEGs).

\subsection{Comparison with previous works}

Three objects (3CR~79, 3CR~223 and 3CR~234) are in common with the
small sample of detected sources in SWW00. The estimated extinctions
reported by the above authors are consistent with our estimates of
$A_{\rm V}$, but much smaller than $A_{\rm V,torus}$.
\citet{taylor96} studied a heterogeneous sample of RGs and RL QSOs
(matched in total radio luminosity and redshift) using $K$-band
imaging. If the RGs and the RL QSOs have the same intrinsic NIR
luminosity (as postulated by the unified scheme), the observed
difference in their nuclear NIR luminosities could be accounted for by
extinctions $A_{\rm V} \sim$8-26~mag, with an uncertainty of
$\sim$7~mag, consistent with what found in our work. 
\citet{haas04} have studied the MIR and FIR properties of 3CR RGs and
quasars, inferring that HEGs are consistent with being misaligned
BLRGs if their nuclei are obscured by extinctions in the MIR of
$\sim$1-2~mag. Using the standard extinction law of \citet{draine89},
this range corresponds to $A_{\rm V}\sim$27-56, consistent with the
range we found.

It is worth also noticing that a search for nuclear point-like UV
emission in RGs from HST-STIS images (NUV-MAMA detector at
$\sim$2300~\AA) has revealed positive detection only for one HEG,
3CR~198, implying an extinction of $A_{\rm V}$=0.08 mag
\citep{chiaberge02}, perfectly consistent with the value estimated in
this work (-0.12$\pm$0.47 mag).  None of the other HEGs targeted
(3CR285, 3CR~321 and 3CR~452) show point-like UV emission
\citep{chiaberge02}. For the only object with an optical detection
(3CR~285), assuming $A_{\rm UV}$=2.26\, $A_{\rm V}$
\citep{cardelli89}, an optical-UV spectral index 1.0$\pm$0.3
(\citealt{chiaberge02}, consistent with that of RL QSOs,
\citealt{elvis94}) and a UV flux limit of
$\sim$1.9$\times$10$^{-19}$~\ergscma (the lowest detection in
\citealt{chiaberge02}), we estimate an extinction $A_{\rm V}
\ga$3.5~mag, consistent with our previous finding.

\subsection{Obscuration vs orientation ?}

A key ingredient that should be added into this analysis is the source
orientation. In the simplest unified scheme, the system is supposed to
be axisymmetric, with the jet direction corresponding to the symmetry
axis. The dusty torus, perpendicular to it, would hide the central
quasar for angles with the line-of-sight larger than the torus opening
angle.

In principle an estimate of the inclination could be derived from the
observed correlation between the radio core luminosity at 5~GHz and
the radio total luminosity at 408~MHz \citep{giovannini01}, e.g. by
computing the normalized core power $P_{\rm CN}$, defined as the
source core power normalized to that corresponding to an average
observing angle of 60$^{\circ}$ for a given total power
\citep{deruiter90}.  Within the simplest scheme, one would then expect
an anti-correlation between $P_{\rm CN}$ and $A_{\rm V,~torus}$.
Although there seems to be a broad trend for HEGs with lower $P_{\rm
CN}$ to have higher extinction (the trend is stronger if BLRGs are
also considered), the large uncertainties and low statistics do not
allow to draw any significant conclusion.


\section{Discussion}
\label{discussion}

From the above results, the following picture emerges for HEGs.  On
one hand, based on their extended and isotropic properties, HEGs are
consistent with being drawn from the same parent population of BLRGs.
In particular, HEGs and BLRGs have the same distributions of 
$L_{\rm 178}$ and $L_{\rm [OIII]}$. If the narrow-line emission arises 
from photo-ionization by the nuclear radiation field, the similarity in
$L_{\rm [OIII]}$ implies that the intrinsic nuclear luminosities of
these populations should also be similar.  On the other hand, based on
their nuclear optical and NIR luminosities, HEGs and BLRGs do not
appear to be taken from the same parent population, with HEGs being
significantly underluminous. Except for a handful of sources (3CR~18,
3CR~197.1, 3CR~198, 3CR~223.1, 3CR~285 and 3CR~349), obscuration alone cannot
account for such differences.  Moreover, for the great majority of
HEGs the observed nuclear NIR luminosity is insufficient to account
for the narrow-line emission.

However, these findings can be reconciled with the unification
scenario for HEGs and BLRGs if a scattered component is present.
Indeed CCC02 have suggested that the optical nuclear emission in HEGs
can actually be identified with a compact scattering region. It is
thus possible that for the majority of HEGs the observed nuclear
optical emission is dominated by the fraction of the nuclear light
scattered toward the line-of-sight.  At NIR wavelengths, instead, both
a transmitted (direct) and the scattered component contribute to the
observed luminosity. In this (simple) interpretation the fraction of
scattered nuclear light ranges from a few tenths of to a few percent,
while the torus extinction $A_{\rm V,~torus}$ spans $\sim$15 to
50~mag, with only a few exceptions.

This scenario is consistent with independent findings on the best 
studied object of
our sample: 3CR~234.  3CR~234 is characterized by a very high degree
of polarization ($\sim$17\% after starlight subtraction) of both the
optical continuum and the broad H$\alpha$ line \citep{antonucci84},
which can be explained only with polarization induced by scattering;
also the $K$-band continuum is polarized, at the lower level of
$\sim$4.6$\pm$0.4\% (\citealt{sitko91}; see also \citealt{young98}).
The flux and polarization properties at optical and NIR wavelengths
have been modeled by \citet{young98}, who concluded that the observed
light comprises a scattered component and a dichroic view through the
torus which becomes important only at the longer wavelengths. This is
qualitatively consistent with our results, although the nuclear
extinction they estimated ($\sim$60 mag) is larger than ours
(22$\pm$2~mag).

Among the sample of HEGs we found 6 sources requiring low 
obscuration. Four of them (3CR~18, 3CR~197.1, 3CR~198, 3CR~349) are  
consistent with $A_{\rm V,torus} <$2~mag, i.e. essentially no 
obscuration, apparently in contrast with the lack of broad lines 
in their optical spectra. Interestingly, results obtained during 
the preparation of this work indicate that 3CR~18 and 3CR~197.1 
are actually BLRGs, misclassified as HEGs. A high S/N VLT spectrum 
revealed a prominent broad (FWHM$\sim$4500 km~s$^{-1}$) H$\alpha$ 
line in 3CR~18 and a broad H$\alpha$ line is clearly visible in the 
SDSS spectrum of 3CR~197.1. 
The hypothesis that misclassification affects also 3CR~198 and 3CR~349 
appears more likely than the alternative explanation that they actually 
lack a BLR. Conversely, 3CR~223.1 and 3CR~285 show sufficient obscuration 
to prevent the detection of broad lines in the optical (e.g. the H$\alpha$ 
would be reduced by a factor $>$ 40), but should be easily detected in 
the NIR. The presence of these low-extinction objects among the sample 
is particularly interesting with respect to whether there is 
continuity between HEGs and BLRGs -- i.e. if the torus optical depth
varies smoothly or if a minimum threshold exists. Higher quality optical 
and NIR spectra are clearly needed to settle this issue.

The population of HEGs can thus be characterized by a distribution in
the amount of absorbing dusty material, although it is not possible
from the current data to determine whether this is closely associated
to the orientation (i.e. the vertical structure of the torus) or
simply to different sources having different amount of dust.  In any
case, for a handful of HEGs the direct view of the central engine is
possible both at optical and NIR wavelengths (might be because the
line-of-sight intercepts only the edges of the torus). At higher
levels of extinction the nuclear optical light from the central engine
is absorbed. The observed point-like optical emission is due 
to a compact scattering region located `outside' the obscuring torus.
At NIR wavelengths, instead, the central nucleus is still directly
visible, and both the direct/transmitted light passing through the
torus and the scattered components contribute to the observed NIR
luminosity.  This is the situation for the majority of HEGs. The
possibility that even higher extinction is present is some sources
remains open (e.g. 3C~379.1 ?), although their number is extremely
limited according to our statistics (note that for very high
extinction also radiation re-processed by the dust could contribute to
the observed emission).

Let us finally compare the ranges of values of the scattering fraction
and obscuration we inferred with respect to those estimated in the
literature for obscured AGN (also non RL, namely Seyfert 2 galaxies).
The scattered fraction in the latter ones spans from less than
$\sim$0.1\% (e.g.  NGC~1068, \citealt{miller91}; 3CR~321,
\citealt{young96a}, 1996b) up to a few \% (e.g. \citealt{diserego94};
\citealt{kishimoto02a}; \citealt{maiolino00}).  A rather conservative
upper limit of $f<$ 10\% has been estimated for the Seyfert~2 Mrk~477
\citep{kishimoto02b}. Our values of $f$ are thus broadly consistent
with those reported for Seyfert~2, although only for a very few
sources this could be truly determined.  Also the inferred values for
the torus extinction appear to be consistent with those
observationally and theoretically estimated for Seyfert~2.  For
example, some torus models (e.g.  \citealt{granato97};
\citealt{fadda98}) require -- in order to reproduce their IR spectra
-- moderate-to-high visual extinctions ($A_{\rm V}$=10-80~mag). The
detection of NIR broad emission lines in some Seyfert 2 (e.g.
\citealt{veilleux97}) also indicates that at these wavelengths the
nuclei can be directly seen, with estimated $A_{\rm V}$ from
$\sim$8~mag up to more than 68~mag.


\section{Conclusions}
\label{conclusion}

We have presented $K^{\prime}$-band observations of all 3CR HEGs
(FR~II) at $z <$0.3, providing for the first time homogeneous NIR
imaging of a complete sub-sample of RGs.  After showing that the
technique based on the surface brightness decomposition to measure
central point-like sources is affected by large uncertainties for the
objects in the studied sample, we have introduced a more accurate
method exploiting the $R-K^{\prime}$ color profiles to measure the NIR
luminosity of a central point-like source.  In all but two HEGs, a
substantial $K^{\prime}$-band excess is found, most likely associated
to the nuclear emission.

The NIR (2.12~$\mu$m) nuclear luminosities have been measured for the
sources, and used to test the unification scheme for RL AGNs.  The
picture emerging from this work is that for the majority of the
studied HEGs, obscuration alone cannot account for the different
nuclear properties of HEGs and BLRGs, and also scattering of the
(optically) hidden nuclear light toward the line-of-sight should play 
a role. For $\sim$70\% of the HEGs, the observed point-like optical
emission is dominated by such scattered component (the direct one
being totally absorbed), while in the $K^{\prime}$-band both
transmitted light and the scattered one contribute to the observed
nuclear luminosity. The estimated fraction of scattered light ranges
from a few tenths of to a few percent of the hidden AGN light, while the
torus extinction is in the range 15$<A_{\rm V,~torus}<$50~mag, with
only a few cases with lower obscuration.  Interestingly, two of the
four HEGs consistent with low levels of obscuration 
($A_{\rm V,~torus}<2$ mag) turned out to be mis-classified objects in 
which a prominent BLR emerges in high quality spectra. This hints to the
possibility that the remaining two low-obscuration objects are also
mis-classified BLRGs. 

To further test this scenario, deep NIR spectropolarimetric
observations are fundamental for two reasons.  Firstly, they should
allow to detect hidden broad lines even when affected by a certain
amount of extinction. Secondly, a measurement of their polarization
can give an estimate of the scattering fraction.  Another important
piece of information could be provided by polarimetric,
multiwavelength observations, as they would allow to discriminate
between electron and dichroic scattering.  HST-NICMOS observations
will be clearly of key importance to determine the presence and
measure the luminosity of nuclear point-like sources.  Finally, X-ray
spectroscopy can provide estimates of the nuclear extinction: assuming
a standard gas-to-dust ratio ($A_{\rm V}=5\times10^{-22} N_{\rm H}$,
\citealt{bohlin78}), we expect the HEGs in our sample to be
Compton-thin, with column densities in the range 
$N_{\rm  H}$=0.3-1$\times$10$^{23}$~cm$^{-2}$.

\begin{acknowledgements}
DM and AC acknowledge the Italian MIUR and INAF for financial
support. We thank M. Chiaberge for useful discussions and the
anonymous referee for comments and suggestions which helped improving
the paper. Funding for the creation and distribution of the SDSS
Archive ({\ttfamily http://www.sdss.org/}) has been provided by the
Alfred P. Sloan Foundation, the Participating Institutions, NASA, NSF,
the U.S.  Department of Energy, the Japanese Monbukagakusho, and the
Max Planck Society.
\end{acknowledgements}



\appendix

\section{Surface brightness decomposition (GALFIT)} \label{galfitApdx}

We have applied the `standard' surface brightness decomposition
technique to detect and measure point sources at the centre of
galaxies by using the two-dimensional fitting algorithm GALFIT
\citep{peng02}. The host galaxy surface brightness profile was modeled
with a \citet{sersic68} profile; the nuclear point-like component was
modeled with a Moffat function, the parameters of which have been
constrained using the stars present in the FoV.

\begin{figure}
\resizebox{\hsize}{!}{\includegraphics{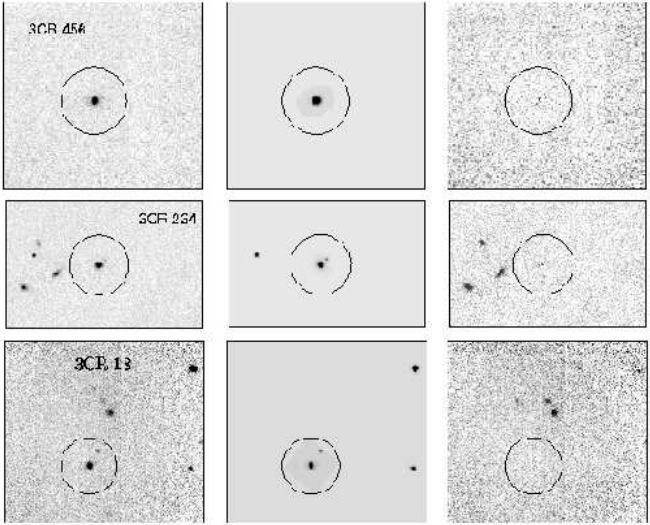}}
\caption{GALFIT imaging decomposition for 3CR~456, 3CR~234 and 
  3CR~18. The left-hand panel is the observed field; the central one
  is the model fitted by GALFIT; the right-hand panel is the residual
  image. The circles indicate the fitted sources.}
\label{Lk_nukeHL}
\end{figure}

In Fig.~\ref{Lk_nukeHL} the results of the GALFIT analysis are shown
for 3CR~456, 3CR~234 and 3CR~18. These three objects have the largest
HST optical nuclear luminosities, $\nu L_{\rm O} >$10$^{43}$~\ergs.
Moreover, 3CR~456 (as detailed in Appendix~\ref{galnotes}) has been
observed with HST-NICMOS, which clearly reveals the presence of a
central point-like source with luminosity
$\sim$10$^{43.63\pm0.09}$~\ergs~at 2.05~$\mu$m (F205W filter).  The
GALFIT best fit solution for 3CR~456 returns a magnitude for the
central point-like source $m_{\rm K^{\prime}}$=14.90$\pm$0.07, which
translates into a nuclear luminosity of 10$^{44.06\pm0.03}$~\ergs~at
2.12~$\mu$m. This is $\sim$0.6~dex larger than the value from the HST
image.  For 3CR~234, the GALFIT analysis gives 
$m_{\rm K^{\prime}}$=13.86$\pm$0.07, implying a luminosity of
10$^{44.25\pm0.03}$~\ergs. This value is consistent with the one
measured by SWW00 of $\sim$10$^{44.32}$~\ergs~and by
\citet{taylor96} of $\sim$10$^{44.2}$~\ergs~from ground-based $K$-band
observations. As detailed in Appendix~\ref{galnotes}, the value
measured with GALFIT is also consistent with that obtained with the
color profile technique.  Finally for 3CR~18, GALFIT detects a central
point-like source with magnitude $m_{\rm K^{\prime}}$=14.97$\pm$0.07,
i.e. a nuclear luminosity of $\sim$10$^{43.83\pm0.03}$~\ergs. As for
3CR~234, this value is consistent within a factor $\sim$1.5 with the
nuclear luminosity measured via the color profile.

\begin{figure}
\resizebox{\hsize}{!}{\includegraphics{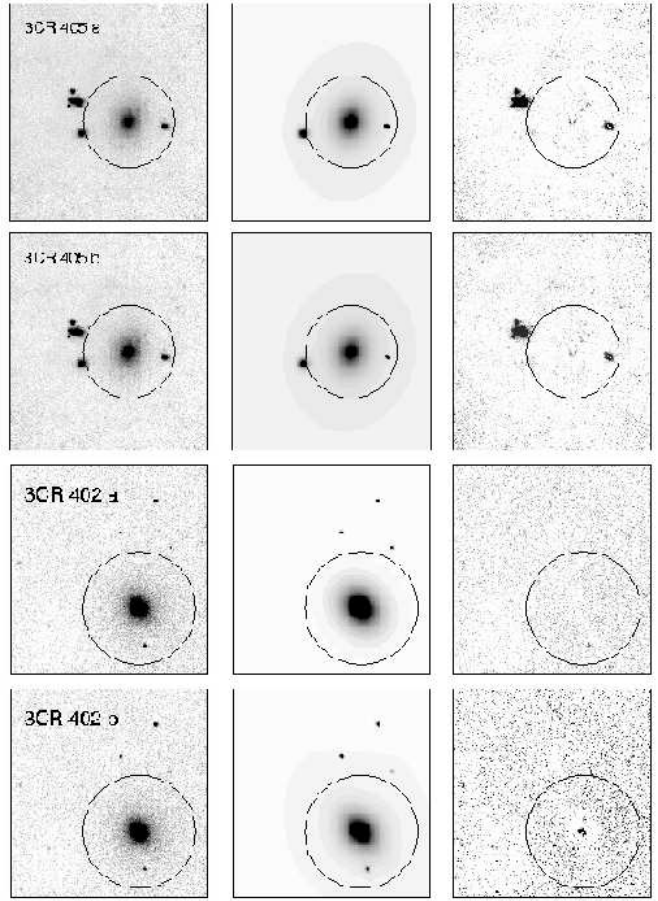}}
\caption{GALFIT imaging decomposition for 3CR~405 and 3CR~402. 
  Panels as in Fig.~\ref{Lk_nukeHL}.  Case A does not include in the 
  modeling analysis a central point-like source, while case B does.}
\label{Lk_nukeLL}
\end{figure}

In Fig.~\ref{Lk_nukeLL} the results from the GALFIT analysis for
3CR~402 and 3CR~405 are shown. These objects have been chosen because
the former is the nearest source ($z$=0.025), while 3CR~405 has been
observed with NICMOS \citep{tadhunter99}. For both sources, we
performed the GALFIT analysis without and with a central point-like
source (case A and case B, respectively).  For 3CR~405, the NICMOS
image (F222M filter) clearly shows a point-like source with a nuclear
luminosity of (4.1$\pm$0.2)$\times$10$^{41}$~\ergs~at 2.25~$\mu$m
\citep{tadhunter99}. The GALFIT analysis performed without adding a
point-like source is identical to that with a central source with
nuclear luminosity a factor $\sim$2.5 larger than the NICMOS one:
based on the goodness of the fit, it is impossible to detect the
presence of a nuclear source in the ground-based image.  Moreover, the
best-fit brightness of the point-like source measured with GALFIT
depends on the initial guessed value, with differences as large as
3.2~mag between best-fit models with the same goodness.  A similar
result is obtained for 3CR~402, for which the quality of the best-fit
models with and without a central point-like source is the same.

We conclude that the standard technique to detect and measure
point-like sources at the centre of galaxies (performed with GALFIT)
returns stable and consistent results only when the point-like source
dominates over the host galaxy (as for 3CR~456 and 3CR~234).  This is
not the case for the great majority of the objects in the studied
sample.  As shown for 3CR~405 and 3CR~402, the standard technique
cannot be used for faint point-like sources, even for the nearest
objects, as the GALFIT analysis becomes unable to detect such
component and the resulting luminosities become strongly dependent on
the initial values.


\section{The $R-K^{\prime}$ color profile technique} \label{colprofApdx}

The $R-K^{\prime}$ color profile technique is here described in
detail. We consider two representative sources. For 3CR~403, a fair
example of the `good quality' sample, we illustrate the whole
technique, i.e.: estimate the nuclear IR luminosity in the
hypothesis of a `well-behaved' color profile, namely a constant outer
(galaxy) $R-K^{\prime}$ profile; show how the uncertainty on the
seeing measurement affects the final result; repeat the analysis in
the more realistic case in which the host galaxy shows color
gradients.  We then consider 3CR~300, representative of the `poor
quality' sample, having a rather noisy color profile.

\subsection{3CR~403} \label{colprofApdx_403}

As explained in \S~\ref{method}, the basic idea is that the presence
of an obscured nucleus will reveal itself with an increase of the IR
nuclear flux with respect to what is seen in the optical images. In
order to compare the NIR and optical images, it is necessary to
produce a synthetic HST image that matches both the seeing conditions
and the pixel size of each IR image.  We fitted a Gaussian to all
objects in the HST FoV which are clearly stellar and measured their
FWHM. The adopted FWHM is their median value and its uncertainty was
estimated from their dispersion. For 3CR~403, FWHM=0\farcs89$\pm$0.08,
and therefore the error on the seeing is $\sim$9\% -- slightly larger 
than that typical for the studied sample. Each HST image has been then 
convolved with a Gaussian with the appropriate FWHM and interpolated 
to match the pixel size of the TNG image. On the `matched' HST image and 
on the reduced TNG one we performed photometry on annuli of increasing 
radius out to 3\arcsec and an average color (and its associated error) 
was derived for each annulus, providing us with the radial color dependence 
of the source.

\begin{figure}
  \resizebox{\hsize}{!}{\includegraphics{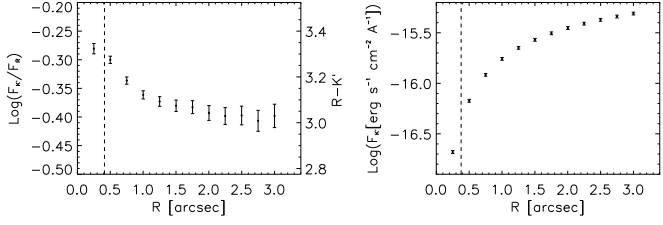}}
\caption{$R-K^{\prime}$ color radial profile (left-hand panel) 
  and aperture flux radial profile (right-hand panel) for 3CR~403. The
  dashed vertical line represents the dispersion (FWHM divided by
  2.354) of the point-spread function of the image. The error bars
  indicate 1$\sigma$ errors.}
\label{color3C403_1}
\end{figure}

The $R-K^{\prime}$ color radial profile, shown in Fig~\ref{color3C403_1}
(left-hand panel), is obtained from:
\begin{equation} 
R-K^{\prime} = 2.5 \log{\Bigg(\frac{F_{\rm K^{\prime},an}}{F_{\rm R,an}}\Bigg)} + 4.025,
\end{equation}
where $F_{\rm K^{\prime},an}$ and $F_{\rm R,an}$ are the annulus
fluxes in the TNG and HST image, respectively; the corresponding error
is estimated with the standard error propagation.

To estimate the nuclear monochromatic $K^{\prime}$-band luminosity 
$L_{\rm K^{\prime}}$, the $K^{\prime}$-band aperture flux
($F_{\rm K^{\prime}}$) radial profile is also needed and estimated
from the counts within the circular aperture (see
Fig.~\ref{color3C403_1}, right-hand panel).

For a `well-behaved' $R-K^{\prime}$ profile, $L_{\rm K^{\prime}}$ can
be then estimated with the following steps:

\begin{enumerate}
\item the reference host galaxy `color', $\left.F_{\rm
K^{\prime}}/F_{\rm R}\right|_{\rm gal}$, is measured in the outer part
of the profile (at $r$=2\farcs75 for 3CR~403);
\item the aperture-corrected nuclear IR excess luminosity $L_{\rm
K^{\prime}, xs}$ is obtained by multiplying the measured nuclear IR
excess [defined as ${1- [(\left.F_{\rm K^{\prime}}/F_{\rm
R}\right|_{\rm gal})/(F_{\rm K^{\prime}}/F_{\rm R})]}$, where $F_{\rm
K^{\prime}}/F_{\rm R}$ is the nuclear color] with the corresponding 
$F_{\rm K^{\prime}}$.
\item the HST optical nuclear component $L_{\rm O}$ is properly scaled
with the host galaxy color $\left.F_{\rm K^{\prime}}/F_{\rm
R}\right|_{\rm gal}$.
\item $L_{\rm K^{\prime}}$ is finally estimated as 
$L_{\rm K^{\prime}}=L_{\rm K^{\prime},scaled}+L_{\rm K^{\prime},xs}$. The
error on $L_{\rm K^{\prime}}$ has been inferred from the standard
error propagation, taking into account also the error on $L_{\rm O}$
($\sim$10-20\%).
\end{enumerate}

The above procedure has been repeated twice to take into account the
uncertainty on the seeing determination: the HST image has been
convolved with a Gaussian with a FWHM increased and decreased by its
1$\sigma$ uncertainty and the whole analysis repeated.  In
Table~\ref{tab3} the derived luminosities are listed. The difference
in the final $L_{\rm K^{\prime}}$ for the different seeing is within
the 1$\sigma$ error and of order $\sim$0.07~dex.

\begin{table}
\caption{IR luminosities for 3CR~403: fixed galaxy color}
\small
\begin{tabular}{l | c c c} 
\hline
FWHM  & $\log{L_{\rm K^{\prime},xs}}$ & $\log{L_{\rm K^{\prime},scaled}}$ & 
$\log{\nu L_{\rm K^{\prime}}}$ \\
($^{\prime \prime}$) & (\ergshz) & (\ergshz) & (\ergs) \\
\hline
0\farcs89             & 28.28$\pm$0.07 & 27.21$\pm$0.07 & 42.47$\pm$0.07 \\
0\farcs89$+$1$\sigma$ & 28.34$\pm$0.06 & 27.21$\pm$0.07 & 42.53$\pm$0.06 \\
0\farcs89$-$1$\sigma$ & 28.21$\pm$0.09 & 27.21$\pm$0.07 & 42.40$\pm$0.08 \\
\hline
\end{tabular}
\label{tab3}
\end{table}

\begin{figure}
\resizebox{\hsize}{!}{\includegraphics{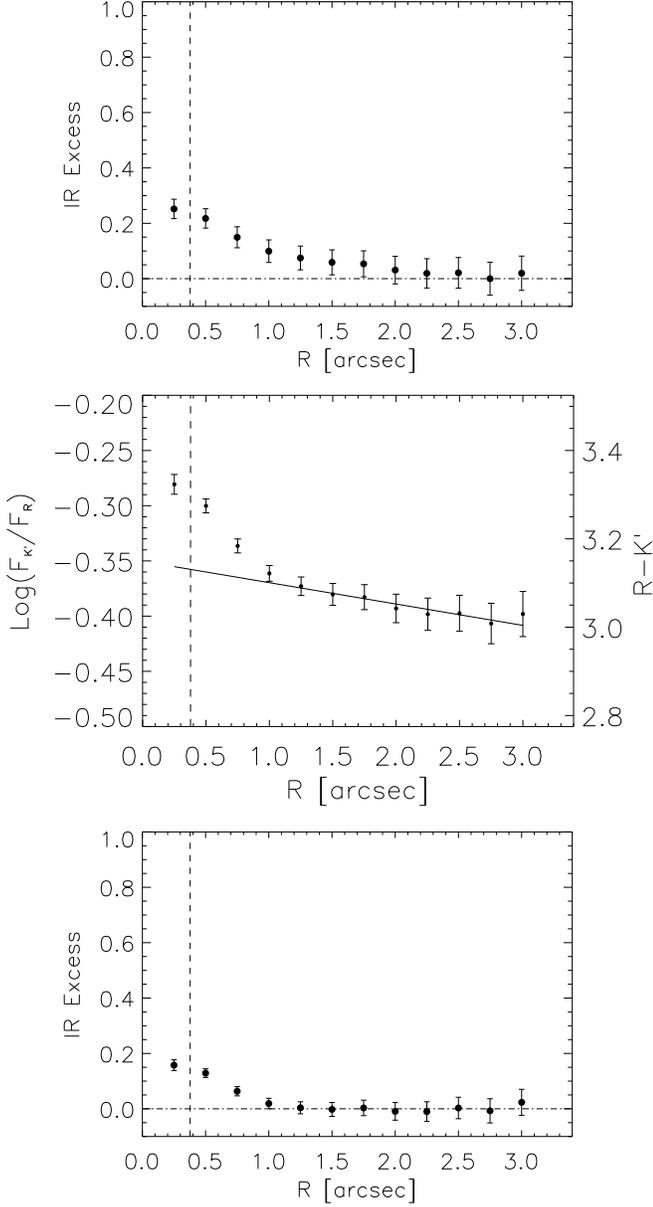}}
\caption{Upper panel: the radial profile of the IR 
  excess 
  for 3CR~403 in the hypothesis of a constant host
  galaxy color; at radii larger than 1\farcs25 a small diffuse IR
  excess indicates that the host color profile is not perfectly flat.
  Middle panel: the observed color radial profile with over plotted
  the best-fit model to the color gradient at radii larger than
  1\farcs25. Lower panel: as the upper panel after modeling the small
  (but significant) color gradient: only the nuclear IR excess is
  present (while the diffuse one is no more present).  The vertical
  dashed line is as in Fig.~\ref{color3C403_1}; the dot-dashed line
  represents the zero IR excess line. Error bars are 1$\sigma$
  errors.}
\label{color3C403_2}
\end{figure}

As shown in Fig.~\ref{color3C403_2} (upper panel), there is a diffuse
IR excess at radii $\ga$1\farcs25 due to a small color gradient,
neglected so far. Such gradient has been fitted with a linear model
$\left.R-K^{\prime}\right|_{\rm model}=A+Br$ at radii $\ga$1\farcs25,
and extrapolated at smaller radii [the profile and the fitted model
are shown in Fig.~\ref{color3C403_2} (middle panel)]. This best-fit
model is then used as the host galaxy reference $\left.F_{\rm
K^{\prime}}/F_{\rm R}\right|_{\rm gal}$ at each radius
(the associated error has been taken as the scatter around the
best-fit) to derive the effective IR excess (Fig.~\ref{color3C403_2},
lower panel).  In Table~\ref{tab4} the estimated luminosities are
listed, also showing the effects of the seeing uncertainty, which is
now of order $\sim$0.15~dex.

The color gradients indeed represent the main limiting factor for the
accuracy of this method.  However, in most of the cases showing
gradients, the estimates of $L_{\rm K^{\prime}}$ based on a flat
profile or a model of the gradient differ by $\la$ 0.2~dex, and never
more than a factor $\sim$2 (only 3CR~357 and 3CR~379.1 show larger
differences, 0.44 and 0.57~dex respectively).

\begin{table}
\caption{IR luminosities for 3CR~403: modeled galaxy color}
\small
\begin{tabular}{l | c c c} 
\hline
FWHM  & $\log{L_{\rm K^{\prime},xs}}$ & $\log{L_{\rm K^{\prime},scaled}}$ 
& $\log{\nu L_{\rm K^{\prime}}}$ \\
($^{\prime \prime}$) & (\ergshz) & (\ergshz) & (\ergs) \\
\hline
0\farcs89             & 28.05$\pm$0.05 & 27.26$\pm$0.07 & 42.27$\pm$0.05 \\
0\farcs89$+$1$\sigma$ & 28.19$\pm$0.04 & 27.25$\pm$0.07 & 42.39$\pm$0.04 \\
0\farcs89$-$1$\sigma$ & 27.84$\pm$0.09 & 27.27$\pm$0.07 & 42.09$\pm$0.08 \\
\hline
\end{tabular}
\label{tab4}
\end{table}

\subsection{3CR~300}

\begin{figure}
\resizebox{\hsize}{!}{\includegraphics{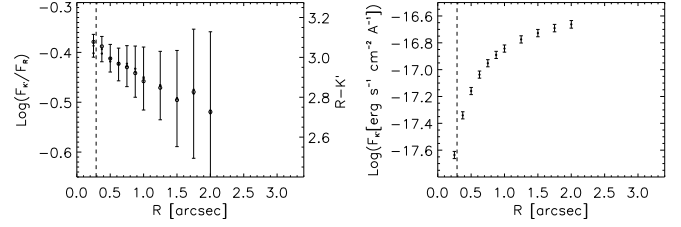}}
\caption{Color profile (left-hand panel) and aperture flux profile 
  (right-hand panel) for 3CR~300. Symbols as in Figs.~\ref{test} and
  \ref{color3C403_1}.}
\label{color3C300_1}
\end{figure}

\begin{figure}
\resizebox{\hsize}{!}{\includegraphics{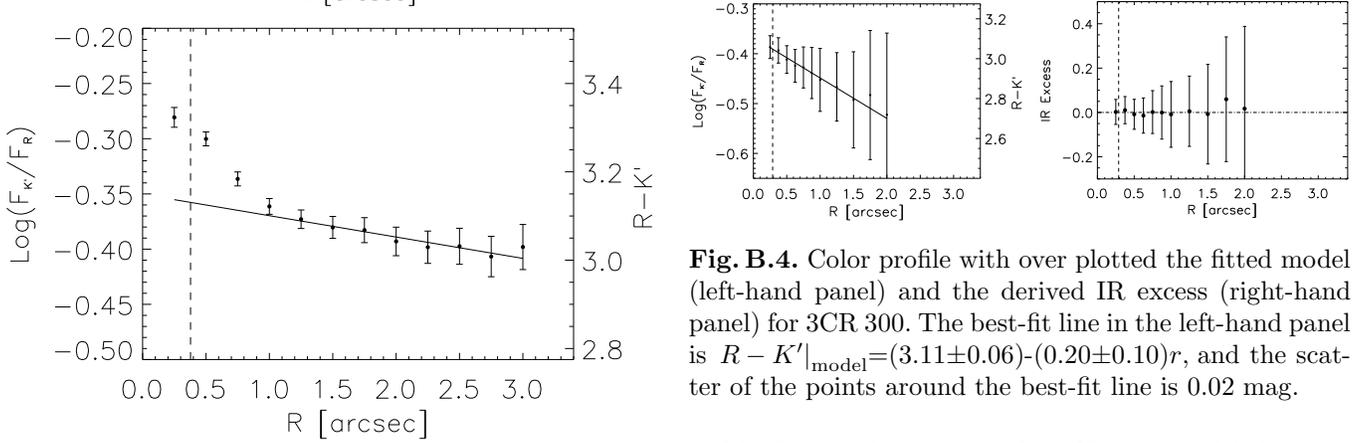}}
\caption{Color profile with over plotted the fitted model (left-hand 
panel) and the derived IR excess (right-hand panel) for 3CR~300. The
best-fit line in the left-hand panel is
$\left.R-K^{\prime}\right|_{\rm model}$=(3.11$\pm$0.06)-(0.20$\pm$0.10)$r$, 
and the scatter of the points around the best-fit line is 0.02~mag.}
\label{color3C300_2}
\end{figure}

In Fig.~\ref{color3C300_1} (left-hand panel), the radial color profile
for 3CR~300 has been plotted (showing also the effects of the seeing
measurement uncertainty, $\sim$2\% for this source). The profile is
quite noisy, and while it is consistent with being constant, a color
gradient seems present. For this reason, we modeled the color profile
as in \S~\ref{colprofApdx_403} (the color profile and the
fitted model are shown in Fig.~\ref{color3C300_2}, left-hand panel).

In Table~\ref{tab5}, the estimated luminosities for 3CR~300 are
reported.  After modeling the color gradient, no significant IR excess
is found (see Fig.~\ref{color3C300_2}, right-hand panel) and
therefore $L_{\rm K^{\prime}} \simeq L_{\rm K^{\prime},scaled}$.

\begin{table}
\caption{IR luminosities for 3CR~300: modeled galaxy color}
\small
\begin{tabular}{l | c c c} 
\hline
FWHM  & $\log{L_{\rm K^{\prime},xs}}$ & $\log{L_{\rm K^{\prime},scaled}}$ 
& $\log{\nu L_{\rm K^{\prime}}}$ \\
($^{\prime \prime}$) & (\ergshz) & (\ergshz) & (\ergs) \\
\hline
0\farcs68             & 27.15$^{+0.84}_{-\infty}$ & 28.36$\pm$0.07 & 42.54$\pm$0.16 \\
0\farcs68$+$1$\sigma$ & 27.26$^{+0.78}_{-\infty}$ & 28.37$\pm$0.07 & 42.55$\pm$0.17 \\
0\farcs68$-$1$\sigma$ & 27.13$^{+0.85}_{-\infty}$ & 28.35$\pm$0.07 & 42.53$\pm$0.16 \\
\hline
\end{tabular}
\label{tab5}
\end{table}


\section{Individual galaxies}\label{galnotes}

Here we report the analysis of the $R-K^{\prime}$ radial profile and
the results used in the estimate of the nuclear IR luminosity for the
individual sources. In Tab.~\ref{tabAppendixLARGE} we listed
the seeing, the IR excess, the reference host galaxy color
$\left.R-K^{\prime}\right|_{\rm gal}$, the parameters of the linear
model fitted to the outer color profile to account for possible
gradients, and short comments. We also comment on individual sources
when needed and compare our results with the previous ones -- based on
the standard technique -- by SWW00 and \citet{taylor96}.

\begin{table*}
\caption{Details of the color profile technique for the individual sources}
\small
\begin{tabular}{l c c c c c l} 
\hline Source & Seeing & IRE & $\left.{R-K^{\prime}}\right|_{\rm gal}$
& $A$ & $B$ & Comments \\ Name & ($^{\prime \prime}$) & (10$^{28}$
erg/s/Hz) & (mag) & (mag) & ($\frac{mag}{^{\prime \prime}}$) & \\ 
& & & & & & \\ \hline 

3C~18 & 1.43$\pm$0.20 & 11.3$\pm$2.9 & 3.51$\pm$0.19 & 3.19$\pm$1.86 & 0.22$\pm$1.12 & \\ 
3C~33 & 0.87$\pm$0.11 & -- & -- & -- & -- & \\
3C~79    & 1.68$\pm$0.03  & 2.40$\pm$2.26     & 3.09$\pm$0.02 & 2.89$\pm$0.29 & 0.17$\pm$0.16  & $K$=14.79$\pm$0.08$^{a}$; $K$=14.79$\pm$0.06 (SWW00)     \\
3C~98    & 1.71$\pm$0.06  & 0.43$\pm$0.03     & 2.93$\pm$0.01 & 2.89$\pm$0.02 & 0.03$\pm$0.01  & $K$=14.71$\pm$0.08; $K$=12.60$\pm$0.03 (SWW00)     \\
3C~105   & 1.73$\pm$0.08  & --                & --	      & --            & --             & \\
3C~153   & 2.23$\pm$0.07  & --                & --	      & --            & --             & \\
3C~171   & 1.02$\pm$0.04  & -9.39$\pm$2.83    & 2.30$\pm$0.14 & --	      & --             & $K$=15.69$\pm$0.12; $K$=15.71$\pm$0.04 (SWW00)     \\
3C~184.1 & 1.26$\pm$0.06  & 3.73$\pm$0.31     & 3.16$\pm$0.05 & 3.23$\pm$0.16 & -0.05$\pm$0.09 & well-behaved color profile \\
3C~197.1 & 0.87$\pm$0.07  & 0.49$\pm$0.29     & 2.87$\pm$0.02 & 2.87$\pm$0.06 & 0.01$\pm$0.05  & well-behaved color profile \\
3C~198   & 0.97$\pm$0.03  & -0.14$\pm$0.40    & 2.62$\pm$0.08 & --            & --             & small nuclear IR deficit \\
3C~223   & 1.36$\pm$0.06  & 1.35$\pm$0.21     & 2.86$\pm$0.03 & 2.72$\pm$0.17 & 0.13$\pm$0.11  & $K$=14.54$\pm$0.08; $K$=14.58$\pm$0.05 (SWW00)     \\
3C~223.1 & 0.73$\pm$0.11  & 7.00$\pm$0.71     & 2.87$\pm$0.03 & --            & --             & well-behaved color profile \\
3C~234   & 0.90$\pm$0.05  & 56.1$\pm$5.4      & 2.31$\pm$0.13 & --            & --             & $K$=13.48$\pm$0.08; $K$=13.49$\pm$0.05 (SWW00)     \\
3C~284   & 0.63$\pm$0.07  & 1.55$\pm$1.19     & 2.99$\pm$0.06 & --            & --             & complex profile with a small bump \\
3C~285   & 0.89$\pm$0.04  & 1.55$\pm$0.13     & --            & 3.13$\pm$0.02 & -0.12$\pm$0.01 & largest outer color gradient \\
3C~300   & 0.68$\pm$0.02  & 0.14$\pm$0.27     & 2.79$\pm$0.24 & 3.11$\pm$0.06 & -0.20$\pm$0.10 & noisy color profile \\
3C~321   & 0.92$\pm$0.13  & 3.68$\pm$0.17     & 2.60$\pm$0.06 & 2.63$\pm$0.17 & -0.01$\pm$0.08 & well-behaved color profile \\
3C~327   & 0.61$\pm$0.06  & $<$0.28           & 3.06$\pm$0.04 & --            & --             & flat color profile at $r \la$1\arcsec \\
3C~349   & 0.81$\pm$0.03  & 0.55$\pm$1.49     & 3.15$\pm$0.10 & 3.38$\pm$0.05 & -0.22$\pm$0.07 & noisy color profile \\
3C~357   & 0.71$\pm$0.10  & 0.76$\pm$0.84     & 2.97$\pm$0.12 & 3.16$\pm$0.03 & -0.09$\pm$0.03 & noisy color profile \\
3C~379.1 & 0.67$\pm$0.03 & $\la$2.16$\pm$6.57 & 3.22$\pm$0.21 & 3.36$\pm$0.09 & -0.12$\pm$0.10 & very noisy color profile \\
3C~381   & 0.80$\pm$0.11  & 5.40$\pm$1.03     & 3.12$\pm$0.15 & 3.10$\pm$0.40 & 0.01$\pm$0.27  & well-behaved color profile \\
3C~402   & 0.77$\pm$0.08  & 0.15$\pm$0.03     & 2.82$\pm$0.02 & --            & --             & well-behaved color profile \\
3C~403   & 0.89$\pm$0.08  & 1.13$\pm$0.14     & 3.01$\pm$0.05 & 3.15$\pm$0.04 & -0.05$\pm$0.02 & well-behaved color profile \\
3C~405   & 0.95$\pm$0.04  & 0.32$\pm$0.07     & 3.29$\pm$0.03 & 3.46$\pm$0.02 & -0.06$\pm$0.01 & well-behaved color profile \\
3C~436   & 0.72$\pm$0.20  & 2.37$\pm$1.94     & 3.33$\pm$0.23 & 3.61$\pm$0.10 & -0.20$\pm$0.09 & noisy profile \\
3C~452   & 0.96$\pm$0.04  & 0.46$\pm$0.17     & 3.04$\pm$0.04 & 3.14$\pm$0.07 & -0.05$\pm$0.04 & well-behaved color profile \\
3C~456   & 1.00$\pm$0.07  & 10.8$\pm$1.0      & 2.92$\pm$0.07 & 2.64$\pm$0.41 & 0.28$\pm$0.36  & $K$=14.64$\pm$0.06; $K$=14.63$\pm$0.10 (dV98) \\
3C~460   & 0.88$\pm$0.10 & $\la$2.20$\pm$6.49 & 3.31$\pm$0.17 & 3.46$\pm$0.31 & -0.04$\pm$0.24 & $K$=14.69$\pm$0.08; $K$=14.71$\pm$0.10 (dV98) \\
\hline
\end{tabular}

{IRE is the nuclear aperture-corrected infrared (2.12~$\mu$m) excess
$L_{\rm K^{\prime},xs}$; $\left.{R-K^{\prime}}\right|_{\rm gal}$ is
the host galaxy color; $A$ and $B$ are the parameters of the linear
model $\left.{R-K^{\prime}}\right|_{\rm model}=A+Br$ fitted to the
observed outer color profiles to account for possible color gradients
[when not specified, the outer profile is consistent with being flat
(3CR~198, 3CR~223.1, 3CR~234, 3CR~402) or complex (3CR~171, 3CR~284)];
$^{a}$ $K$ is the $K-$band photometry within a 3$^{\prime
\prime}$-aperture for the objects from SWW00, and within a
7\farcs5-aperture for the sources from \citet{devries98} (dV98).}

\label{tabAppendixLARGE}

\end{table*}

\noindent 
{\bf 3CR~079}: SWW00 measured $\nu L_{\rm K}=$10$^{43.66}$ \ergs, a 
factor of $\sim$4-6 larger than our measurement, and a factor
3-5 fainter than the one reported in \citet{taylor96}. The
discrepancy between these measurements is not easily
understandable. While the observed flat $R-K^{\prime}$ profile
excludes a possible underestimate of $L_{\rm K^{\prime}}$, the
standard technique requires a more uncertain modeling of the host
galaxy, possibly causing to underestimate the host contribution.

\noindent 
{\bf 3CR~098}: SWW00 measured $\nu L_{\rm K}<$10$^{41.28}$ \ergs,
while we obtain a luminosity a factor of $\sim$3 larger.

\noindent 
{\bf 3CR~171}: A very significant gradient is found, from
$r$=1\farcs75 to the innermost point corresponding to a nuclear IR
deficit of (9.39$\pm$2.83)$\times$10$^{28}$ \ergshz~ at
$r$=0\farcs5. This might be due to an optical excess caused by the
strong line emission present in the inner 4\arcsec~(H$\alpha$ flux of
$\sim$2.12$\times$10$^{-14}$ \ergscm; \citealt{tadhunter00}).

\noindent 
{\bf 3CR~223}: SWW00 measured $\nu L_{\rm K} \sim$ 10$^{43.50}$ \ergs,
a factor of $\sim$10 larger than ours.

\noindent 
{\bf 3CR~234}: Our 3\arcsec~aperture photometry is about 0.4~mag
larger than the value reported in SWW00. However, using their best fit
parameters for the host+nucleus decomposition, we estimate a
3\arcsec~aperture photometry of $K$=13.49$\pm$0.05, perfectly
consistent with our value. SWW00 measured $\nu L_{\rm K}$=10$^{44.32}$
\ergs, a factor $\sim$1.8 larger than our measurement, while
\citet{taylor96} measured $\nu L_{\rm K}$=10$^{44.2}$ \ergs. These two
values are consistent within the errors, and this is not surprising as
this source is dominated by the nuclear emission, and are also consistent 
with our value.

\noindent 
{\bf 3CR~327}: The $R-K^{\prime}$ profile is characterized by a
gradient at radii $>$1\arcsec~and a constant $R-K^{\prime}$ color
toward the center ($R-K^{\prime}$=3.06$\pm$0.04 applying a weighted
mean of the innermost 8 points). The inner plateau implies a 1$\sigma$
upper limit of the nuclear IRE of 2.77$\times$10$^{27}$ \ergshz.

\noindent 
{\bf 3CR~349}: The color profile is quite noisy, and the sky dominates 
the profile at radii $>$1\farcs75. Assuming a flat outer 
($r \ga$0\farcs75) color profile, we estimated 
$\nu L_{\rm K^{\prime}}=$10$^{43.09\pm0.10}$ \ergs. Although 
very noisy, the color profile shows a gradient in the outer part; we 
therefore studied the extreme case in which no IRE is present by modeling 
the color profile. The resulting $\nu L_{\rm K^{\prime}}=$ is 
10$^{43.02\pm0.11}$ \ergs, consistent with the previous estimate.

\noindent 
{\bf 3CR~357}: Assuming a flat outer color profile, we estimated 
$\nu L_{\rm K^{\prime}}=$ 10$^{42.63\pm0.21}$ \ergs; modeling the 
outer  ($r \geq$0\farcs4) galaxy profile, we obtained 
$\nu L_{\rm K^{\prime}}=$ 10$^{42.19\pm0.34}$ \ergs,
0.44~dex smaller than the former estimate.  

\noindent
This object and 3CR~379.1 are the worst ones, with $\nu L_{\rm K^{\prime}}$
badly constrained (but still within an uncertainty of a factor
$\sim$3).  In the HST image, extended dust lanes are clearly visible
within the galaxy, down to the very inner region. In order to test the
sensitivity of our analysis to extended dust lanes, we modeled the
galaxy in the HST image with an elliptical model (masking out all the
visible dust lanes). Then we repeated our analysis using the
`dust-free' modeled galaxy profile: the two $R-K^{\prime}$ profiles
are very similar (differences well within the error bars).  Therefore,
we conclude that our analysis is not significantly sensitive to
extended dust lanes.

\noindent 
{\bf 3CR~405}: The observed $R-K^{\prime}$ profile is very 
well-behaved, being flat at radii $>$2\arcsec~with variation 
$<$0.01 mag. The nuclear IRE measured is 
(5.13$\pm$0.64)$\times$10$^{27}$ \ergshz. 3CR~405 (a.k.a. Cygnus~A) 
has been previously observed with NICMOS \citep{tadhunter99} 
(F222M filter centered at 2.25 $\mu$m). \citet{tadhunter99} report 
the presence of a nuclear point source (unresolved, with FWHM 
$<$0\farcs21) which dominates the emission at 2.25~$\mu$m with a flux 
$F_{\rm 2.25 \mu m}$=(4.9$\pm$1.0)$\times$10$^{-28}$ \ergscmhz. This 
translates into a luminosity of (4.1$\pm$0.2)$\times$10$^{41}$ 
\ergs~at 2.25 $\mu$m, about a factor $\sim$1.8 smaller than the value 
reported in this work of (7.3$\pm$0.9)$\times$10$^{41}$ \ergs~at 2.12 
$\mu$m within an aperture of $\sim$0\farcs75. Our larger value is due 
to the fact that we did neglect the color gradient at $r \ga$1\farcs25: 
if this gradient is fitted with a linear model we obtain 
$\nu L_{\rm K^{\prime}}$=(4.57$\pm$0.94)$\times$10$^{41}$, perfectly 
consistent with the NICMOS value. 

\noindent 
{\bf 3CR~456}: The F205W filter image from the STScI public archive 
clearly shows the presence of a nuclear point-like source
with $\nu L_{\rm \nu}$=10$^{43.63\pm0.09}$ \ergs~ at 2.05~$\mu$m (the
error, about 20\%, is dominated by the uncertainty on the subtracted
background) consistent at the 1.8$\sigma$ level with our measurement.

\noindent 
{\bf 3CR~460}: 3CR~460 has been observed with NICMOS (F205W filter).
From the NIC2 image in the public archive the emission is
extended, and there is no clear sign of a nuclear point-like
source. We set as an upper limit the light excess of the central
circular aperture of radius 0\farcs17, obtaining $\nu L_{\rm \nu}
<$10$^{42.82\pm0.04}$ \ergs~ at 2.05~$\mu$m, consistent with our
measurement of 10$^{42.79\pm0.19}$ \ergs~ at 2.12~$\mu$m.

\section{Optical and NIR images} \label{figuresingle}

\begin{figure*}
\centering
\includegraphics[width=18cm]{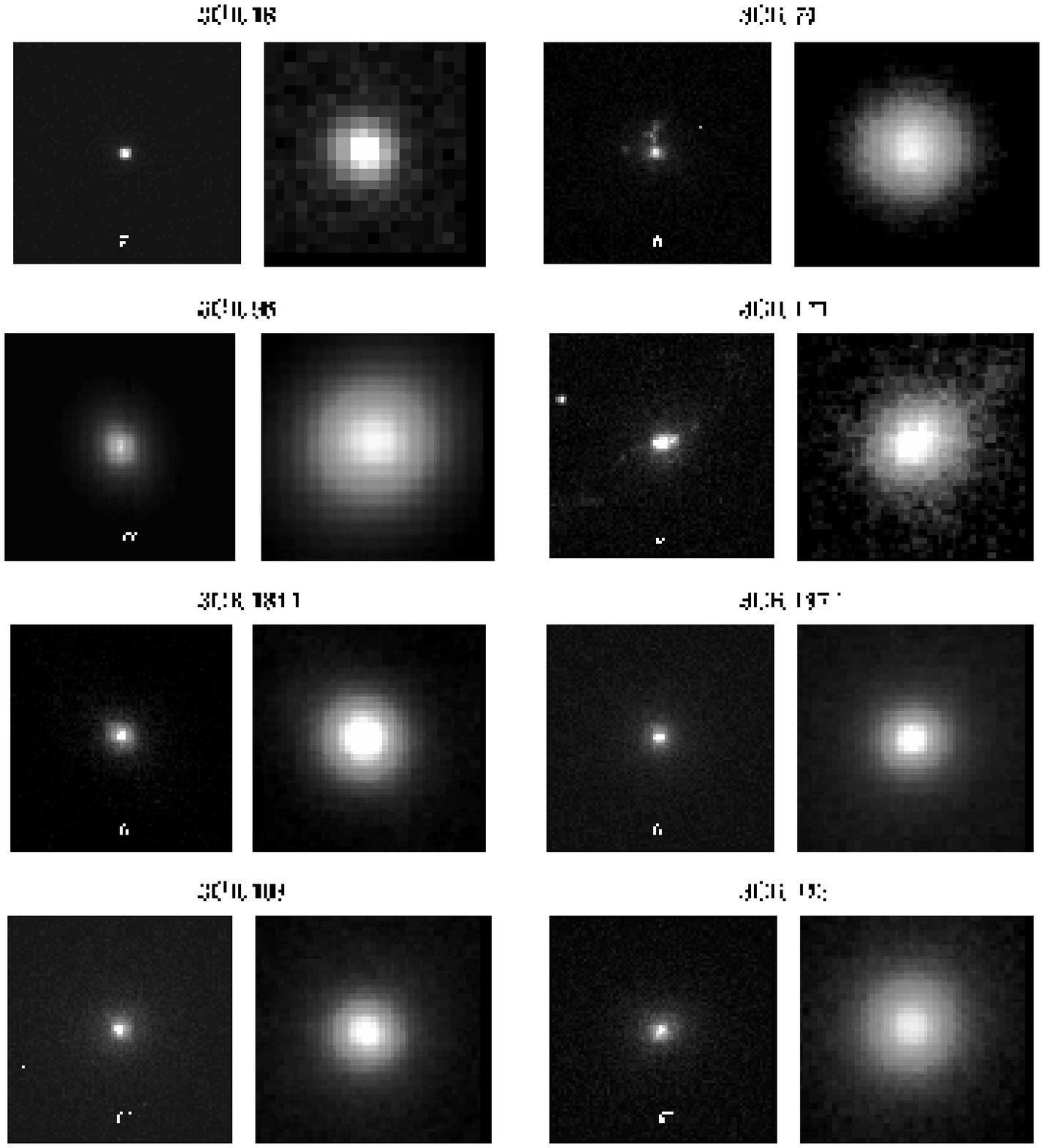}
\caption{HST/WFPC2 F702W images (left-hand panels) and ARNICA/NICS
images (right-hand panels) of the HEG-FRII galaxies of the sample.}
\label{figureIma1}
\end{figure*}

\begin{figure*}
\centering
\includegraphics[width=17cm]{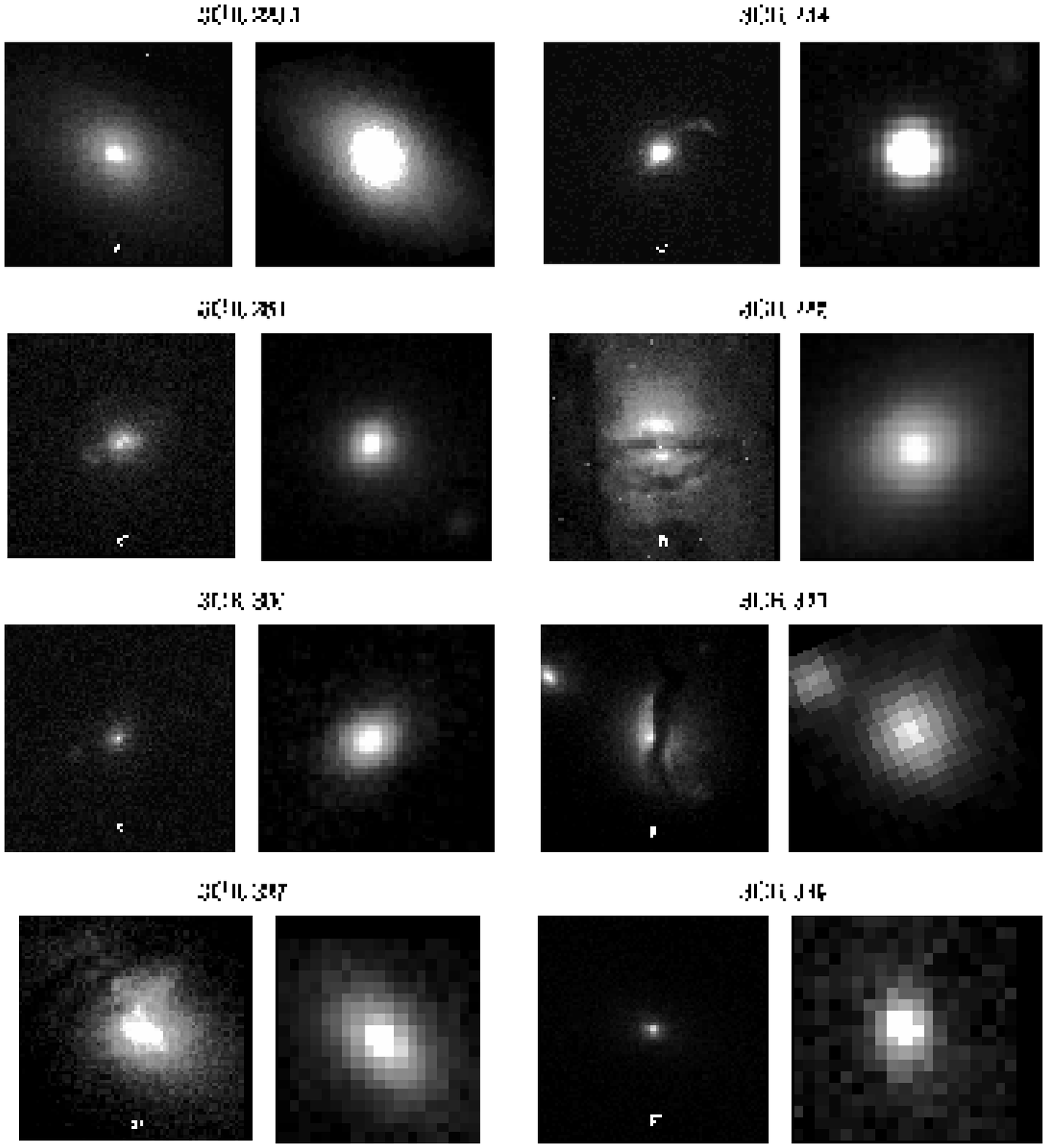}
\caption{(cont.)}
\label{figureIma2}
\end{figure*}

\begin{figure*}
\centering
\includegraphics[width=17cm]{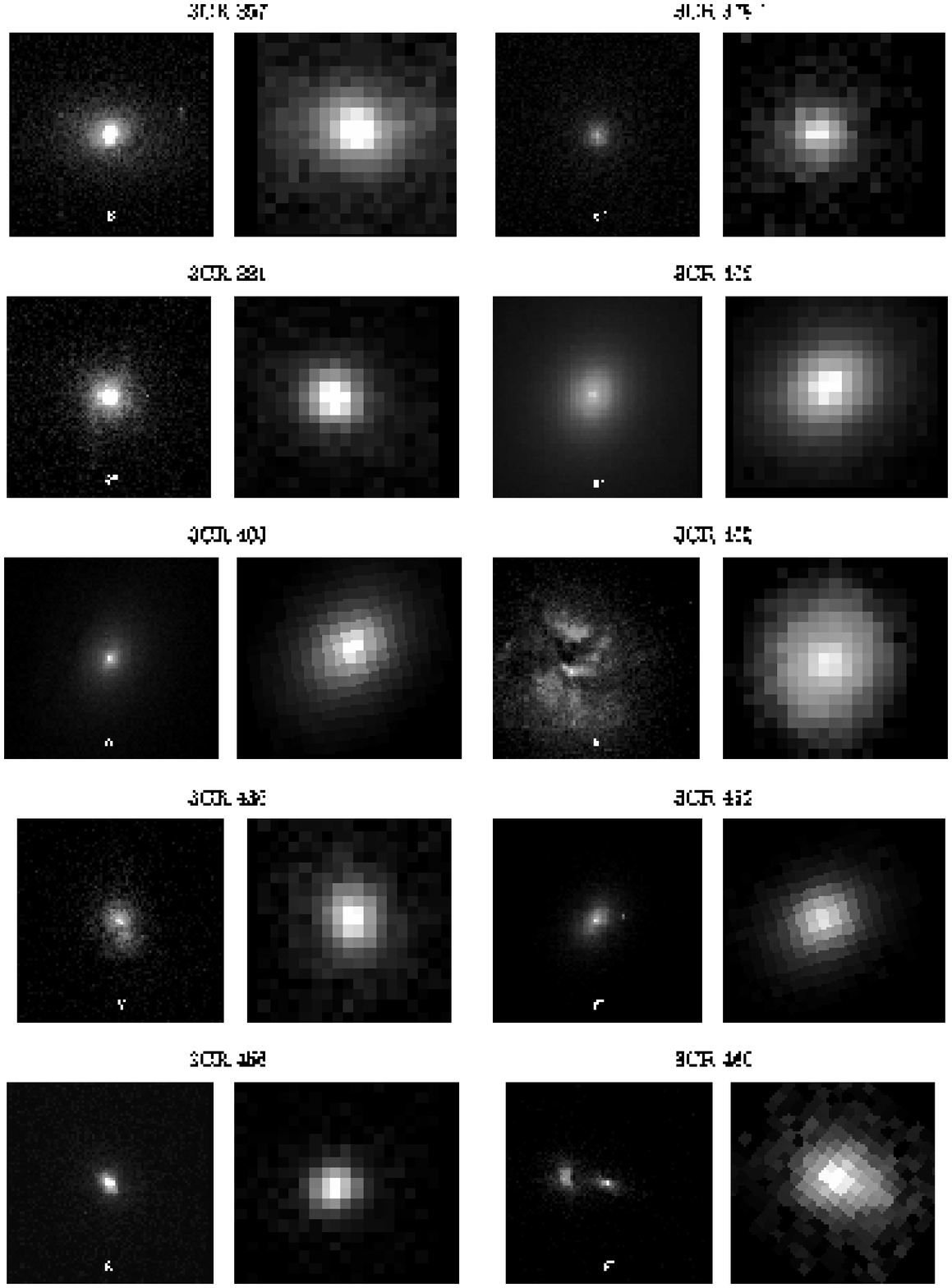}
\caption{(cont.)}
\label{figureIma3}
\end{figure*}

\end{document}